# On the interactions between a propagating shock wave and evaporating water droplets

Zhiwei Huang and Huangwei Zhang*

*Department of Mechanical Engineering, National University of Singapore, 9 Engineering Drive 1, Singapore 117576, Republic of Singapore*

**Abstract**

One-dimensional numerical simulations based on hybrid Eulerian-Lagrangian approach are performed to investigate the interactions between propagating shock waves and dispersed evaporating water droplets in two-phase gas-droplet flows. Two-way coupling for interphase exchanges of mass, momentum and energy is adopted. Parametric study on shock attenuation, droplet evaporation, motion and heating is conducted, through considering various initial droplet diameters (5-20 μm), number densities ($2.5 \times 10^{11}$ - $2 \times 10^{12}$ /$m^3$) and incident shock Mach numbers (1.17-1.9). It is found that the leading shock may be attenuated to sonic wave and even subsonic wave when droplet volume fraction is large and/or incident shock Mach number is low. Attenuation in both strength and propagation speed of the leading shock is mainly caused by momentum transfer to the droplets that interact at the shock front. Total pressure recovery is observed in the evaporation region, whereas pressure loss results from shock compression, droplet drag and pressure gradient force behind the shock front. Recompression of the region between the leading shock and two-phase contact surface is observed when the following compression wave is supersonic. After a critical point, this region gets stable in width and interphase exchanges in mass, momentum, and energy. However, the recompression phenomenon is sensitive to droplet volume fraction and may vanish with high droplet loading. For an incident shock Mach number of 1.6, recompression only occurs when the initial droplet volume fraction is below $3.28 \times 10^{-5}$.

* Corresponding author. E-mail: huangwei.zhang@nus.edu.sg. Tel: +65 6516 2557.

## 1. Introduction

The interactions between propagating shock waves and evaporating droplets are fundamental but challenging multi-phase problems. Studies on such topics have been performed for decades in different fields, e.g. aerospace propulsion[1,2,3], internal combustion engine[4,5,6] and shock tube[7,8,9]. Due to the simplicity in geometry and convenience for parametric study, shock-droplet or shock-particle interactions have been extensively investigated in shock tube facilities. A series of shock tube experiments have been conducted to study the influence of a cloud of water droplets on propagation of a planar shock wave[8]. Although wall pressure data have been recorded for further analysis on shock attenuation, details on droplet evaporation and movement as well as interphase interactions were not attainable due to the limitations of experimental measurement. The interactions of shock wave and a single water droplet with inside vapor cavitation are experimentally studied in a shock tube[9]. The evolutions of both droplet and vapor cavity are recorded with high-speed imaging techniques. However, the focus has been laid on the droplet deformation and cavity collapse. As in other shock tube experiments[10,11], more detailed information is difficult to be measured due to the limitations of measurement techniques, e.g. evolutions of shock strength and Mach number, droplet volume fraction, response timescales and interphase coupling. Therefore, the effects of shock waves on dispersed droplets and the interphase exchanges of mass, momentum and energy are still not well understood in shocked two-phase flows.

A methodology for simulating two-phase flows considering two-way coupling has been developed to investigate the effect of droplet mass and heat transfer on one-dimensional (1D) shock waves[7]. However, the focus was on the shock attenuation, instead of detailed evolutions of droplet properties, e.g. diameter and temperature. Recently, the effect of shock waves on the dispersion characteristics of a particle cloud has been investigated both numerically and analytically[12]. A one-dimensional one-way coupling analytical study is conducted to estimate the cloud topology in the wake of a shock wave. Moreover, two-way formalism is developed, through further considering post-shock gas deceleration due to dispersed particles. However, evaporation, pressure gradient force (PGF), heat transfer and gas viscosity are all neglected in the above work. The only coupling between the two phases is the particle momentum equation with drag force.

In this work, numerical studies are performed with hybrid Eulerian-Lagrangian method to investigate the interactions between a propagating shock wave and evaporating water droplets. The two-way interphase coupling of mass, momentum and energy are considered. To better describe the kinematic effects of shock wave on dispersed droplets, the PGF is considered in our model, besides the drag force. The objectives are two-fold. Firstly, the interactions between dispersed evaporating droplets and shock waves are investigated under a range of operating conditions. Shock attenuation in both strength and Mach number, droplet temperature, velocity and diameter variations will be studied. This differs from the previous work, e.g. in Refs.[7,8,9,10,11], which are mainly focused on shock attenuation or droplet breakup. Secondly, the evolutions of the two-phase interactions are discussed in detail, which allows detailed demonstrations of the unsteady process with novel two-phase flow phenomena, e.g. gas recompression in droplet-laden area. The rest of the paper is organized as follows. Numerical approach, including governing equations and numerical schemes, are described in Section 2. Physical models and mesh sensitivity analysis are given in Sections 3 and 4, respectively. The results and discussion are presented in Section 5, and conclusions are drawn in Section 6.

## 2. Governing equation and numerical method

### *2.1. Governing equations for gas phase*

The governing equations for compressible multi-component flows include the conservation laws of mass, momentum, energy and species mass fraction[13]. They respectively read

$$\frac{\partial \rho_g}{\partial t} + \frac{\partial}{\partial x_i}\left(\rho_g u_{g,i}\right) = S_m, \tag{1}$$

$$\frac{\partial}{\partial t}\left(\rho_g u_{g,i}\right) + \frac{\partial}{\partial x_j}\left(\rho_g u_{g,j} u_{g,i} + p_g \delta_{ij} - \tau_{ij}\right) = S_{M,i}, \tag{2}$$

$$\frac{\partial}{\partial t}\left(\rho_g E_g\right) + \frac{\partial}{\partial x_i}\left(\rho_g E_g u_{g,i} - q_i - \tau_{ij} u_{g,i} + p u_{g,i}\right) = S_e, \tag{3}$$

$$\frac{\partial}{\partial t}\left(\rho_g Y_m\right) + \frac{\partial}{\partial x_i}\left(\rho_g Y_m u_{g,i}\right) - \frac{\partial}{\partial x_i}\left(\rho_g D_g \frac{\partial Y_m}{\partial x_i}\right) = S_{Y_m}, \tag{4}$$

where $t$ is time and $x$ is spatial coordinate. $\rho_g$ is the gas density, $u_{g,i}$ is the gas velocity component, $p_g$ is the gas pressure, $\delta_{ij}$ is Kronecker delta function, and $\tau_{ij}$ is the viscous stress tensor. $E_g$ is the total energy, which is calculated as $E_g = e_{s,g} + \sum_{i=1}^{3} u_{g,i}^2$, with $e_{s,g}$ being the sensible internal energy. $Y_m$ is the mass fraction of $m$-th species, $D_g$ is the molecular diffusion coefficient and $D_g = \mu_g/(\rho_g \cdot Le)$, where $Le$ is Lewis number (assumed to be unity in this work) and $\mu_g$ is the dynamic viscosity. $q_i$ is the $i$-th component of the heat flux $\boldsymbol{q} = -k_g \nabla T_g$, with $k_g$ being the gas thermal conductivity and $T_g$ being gas temperature. The source terms, $S_m$, $S_{M,i}$, $S_e$ and $S_{Ym}$, in Eqs. (1)−(4) denote the exchanges of mass, momentum, energy and species between gas and liquid phases. They are estimated respectively as

$$S_m = -\frac{1}{V_c}\sum_1^{N_{dc}} \dot{m}_d, \tag{5}$$

$$S_{M,i} = -\frac{1}{V_c}\sum_1^{N_{dc}}\left(-\dot{m}_d u_{d,i} + F_{d,i} + F_{p,i}\right), \tag{6}$$

$$S_e = -\frac{1}{V_c}\sum_1^{N_{dc}}\left(\dot{Q}_c + \dot{Q}_{lat}\right), \tag{7}$$

$$S_{Y_m} = \begin{cases} S_m, & \text{for the liquid species,} \\ 0, & \text{for other species.} \end{cases} \tag{8}$$

Here $V_c$ is the volume of a CFD cell, $N_{dc}$ is the droplet number in the cell, $\dot{m}_d$ is the evaporation rate of a single droplet and is given later in Eq. (14), and $u_{d,i}$ is the velocity component of droplet in $i$-th direction. $-\dot{m}_d u_{d,i}$ represents the rate of momentum transfer because of droplet evaporation, whilst $F_{d,i}$ and $F_{p,i}$ are respectively the drag and PGF exerted on the droplet in $i$-th direction and are given in Eqs. (23) and (25). For Eq. (6), other forces (e.g. gravity and Magnus lift force) are not considered in the present work. In Eq. (7), $\dot{Q}_c$ is the convective heat transfer rate (CHTR) between the droplet and gas phases, whilst $\dot{Q}_{lat}$ is the evaporation-induced heat transfer relating to latent heat of droplet vaporization.

*2.2. Governing equations for liquid droplet phase*

The monodispersed liquid phase is modeled as a large number of spherical droplets tracked by Lagrangian method[14]. The interactions between droplets are neglected since dilute spray is studied in this work, in which the volume fraction of dispersed droplets is typically less than 0.1%[15]. The droplet breakup is not considered here since in our simulations the droplet Weber number is generally less than 12. This is lower than the critical Weber number for droplet breakup estimated by Tarnogrodzki[16]. The governing equations of mass, momentum and energy for individual droplets respectively take the following form

$$\frac{dm_d}{dt} = \dot{m}_d, \tag{9}$$

$$\frac{du_{d,i}}{dt} = \frac{F_{d,i}+F_{p,i}}{m_d}, \tag{10}$$

$$c_{p,d}\frac{dT_d}{dt} = \frac{\dot{Q}_c+\dot{Q}_{lat}}{m_d}, \tag{11}$$

where $m_d$ is the mass of a single droplet and can be calculated as $m_d = \pi\rho_d d_d^3/6$ for spherical droplets with $\rho_d$ and $d_d$ being the droplet density and diameter, respectively. $c_{p,d}$ is the droplet heat capacity and $T_d$ is the droplet temperature. Both $\rho_d$ and $c_{p,d}$ are functions of droplet temperature $T_d$ to account for the thermal expansion when droplet is heated[17]

$$\rho_d(T_d) = \frac{a_1}{a_2^{1+(1-T_d/a_3)^{a_4}}}, \tag{12}$$

$$c_{p,d}(T_d) = \frac{b_1^2}{\tau} + b_2 - \tau\left\{2.0b_1b_3 + \tau\left\{b_1b_4 + \tau\left[\frac{1}{3}b_3^2 + \tau\left(\frac{1}{2}b_3b_4 + \frac{1}{5}\tau b_4^2\right)\right]\right\}\right\}, \tag{13}$$

where $a_1$, $a_2$, $a_3$, $a_4$ and $b_1$, $b_2$, $b_3$, $b_4$ are the species-specific constants. $\tau = 1.0 - \tilde{T}/T_{cr}$ with $\tilde{T} = min(T_d, T_{cr})$, where $T_{cr}$ is the critical temperature and $min(\cdot)$ is the minimum function[17].

The evaporation rate, $\dot{m}_d$, in Eq. (9) is estimated through[18]

$$\dot{m}_d = -\pi \rho_s d_d Sh D_{ab} ln(1 + B_M), \tag{14}$$

where $\rho_s = p_s M_d / R_0 T_s$ is the vapor density at the droplet surface. $p_s$ and $T_s$ are respectively the vapor pressure and temperature at the droplet surface, $R_0$ = 8.314 J/(mol·K) is the universal gas constant and $M_d$ is the molecular weight of the vapor. $T_s$ is estimated using the two-thirds rule, i.e. $T_s = (2T_d + T_g)/3$[18]. $B_M$ is the Spalding mass transfer number[18]

$$B_M = \frac{Y_s - Y_g}{1 - Y_s}, \tag{15}$$

where $Y_s$ and $Y_g$ are respectively the vapor mass fractions at the droplet surface and in the ambient gas phase. $Y_s$ can be calculated as

$$Y_s = \frac{M_d X_s}{M_d X_s + M_{ed}(1 - X_s)}, \tag{16}$$

where $M_{ed}$ is the averaged molecular weight of the mixture excluding the vapor at the droplet surface and $X_s$ is the mole fraction of the vapor at the droplet surface, which is calculated using Raoult's Law

$$X_s = X_{liq} \frac{p_{sat}}{p_s}, \tag{17}$$

in which $X_{liq}$ is the mole fraction of the liquid species in the liquid mixture. $p_{sat}$ is the saturated vapor pressure and is estimated as a function of droplet temperature[17]

$$p_{sat}(T_d) = p_g \cdot exp\left(c_1 + \frac{c_2}{T_d} + c_3 ln T_d + c_4 T_d^{c_5}\right), \tag{18}$$

where $c_1$, $c_2$, $c_3$, $c_4$ and $c_5$ are constants[17]. Equation (18) is a modified version of the classical Clasius-Clapeyron equation, and it fits data accurately even much above the normal boiling point[17]. Hence, it is expected to handle the droplet evaporation more accurately in shocked or reactive flows[19]. Similarly, the vapor pressure at the droplet surface $p_s$ is a function of droplet surface temperature $T_s$, i.e.

$$p_s(T_s) = X_{liq} p_g \cdot exp\left(c_1 + \frac{c_2}{T_s} + c_3 ln T_s + c_4 T_s^{c_5}\right). \tag{19}$$

The vapor mass diffusivity in the gaseous mixture in Eq. (14), $D_{ab}$, is modelled as[20]

$$D_{ab} = 10^{-3} \frac{T_s^{1.75}}{p_s} \sqrt{\frac{1}{M_d} + \frac{1}{M_g}} / \left(V_1^{1/3} + V_2^{1/3}\right)^2, \tag{20}$$

where $V_1$ and $V_2$ are constants[21] and $M_g$ is the molecular weight of the carrier gas. Equation (20) has comparable accuracy to the classical Chapman-Enskog kinetic theory, but is more universal, since the latter necessitates empirical parameters and simplifications[21].

The Sherwood number in Eq. (14), *Sh*, is[22]

$$Sh = 2.0 + 0.6Re_d^{1/2}Sc^{1/3}, \tag{21}$$

where *Sc* is the Schmidt number of the gas phase. The droplet Reynolds number in Eq. (21), $Re_d$, is defined based on the slip velocity between two phases, i.e.

$$Re_d \equiv \frac{\rho_g d_d |u_{g,i} - u_{d,i}|}{\mu_g}. \tag{22}$$

The Stokes drag in Eq. (10), $F_{d,i}$, is modeled as (assuming that the droplet is spherical)[23]

$$F_{d,i} = \frac{1}{8}\pi d_d^2 \rho_g C_d |u_{g,i} - u_{d,i}| (u_{g,i} - u_{d,i}), \tag{23}$$

where $C_d$ is the drag coefficient and is estimated as[23]

$$C_d = \begin{cases} \frac{24}{Re_d}\left(1 + \frac{1}{6}Re_d^{2/3}\right), & Re_d \le 1{,}000 \\ 0.424, & Re_d > 1{,}000 \end{cases}. \tag{24}$$

It has been shown from the studies by Cheatham and Kailasanath[24] that the estimations in Eq. (24) can accurately predict the velocity distributions of shock-containing flow fields.

The pressure gradient force in Eq. (10), $F_{p,i}$, accounts for the strong local pressure variation at the rarefaction waves or shock discontinuities. It is

$$F_{p,i} = -\frac{1}{6}\pi d_d^3 \frac{\partial p_g}{\partial x_i}. \tag{25}$$

The convective heat transfer rate $\dot{Q}_c$ in Eq. (11) is

$$\dot{Q}_c = h_c A_d (T_g - T_d), \tag{26}$$

where $A_d = \pi d_d^2$ is the droplet surface area. $h_c$ is the CHT coefficient calculated from the Nusselt number *Nu* using the Ranz and Marshall correlation[22]

$$Nu = \frac{h_c d_d}{k_g} = 2.0 + 0.6Re_d^{1/2}Pr^{1/3}, \tag{27}$$

where $Pr = \mu_g \cdot c_{p,g}/k_g$ is the Prandtl number of the gas and $c_{p,g}$ is the heat capacity at constant pressure.

The latent heat of evaporation, $\dot{Q}_{lat}$, in Eq. (11) is

$$\dot{Q}_{lat} = -\dot{m}_d h(T_d), \tag{28}$$

where $h(T_d)$ is the vapor enthalpy at the droplet temperature $T_d$.

*2.3. Numerical method*

The governing equations of gas and liquid phases are solved by a two-phase multi-component compressible flow solver, *RYrhoCentralFoam*[25]. It is developed from *rhoCentralFoam* solver in OpenFOAM 5.0 package[26]. The cell-centered finite volume method is used to discretize the gas phase equations (Eqs. 1 − 4). Implicit second-order Crank − Nicolson scheme is used for temporal discretization. Second-order central differencing scheme[27] is applied for calculating diffusive fluxes, whereas the semi-discrete Kurganov, Noelle and Petrova (KNP) scheme[28] with Minmod flux limiter[29] is used for convective fluxes. The Lagrangian equations for the droplet phase (Eqs. 9 − 11) are integrated with first-order Euler implicit method. The gas phase quantities at the droplet location, e.g. $p_g$ in Eq. (18), $u_{g,i}$ and $\mu_g$ in Eq. (22), are linearly interpolated from the gas phase results solved from Eqs. (1)−(4). The exchange terms in the gas and droplet equations are updated at each time step with two-way coupling of the two phases. The CFL number of the gas phase equations is 0.02, which corresponds to the physical time step of about $10^{-8}$ s.

The *rhoCentralFoam* solver has been validated by Greenshields *et al.*[26] and Zhang *et al.*[30] with various canonical tests, including Sod's shock tube problem, forward-facing step, supersonic jet and shock-vortex interaction. The results show that the KNP scheme can capture flow discontinuities (e.g. shocks and rarefaction waves) accurately. Furthermore, its accuracy in predicting multi-component gaseous flows has been validated in various reacting or non-reacting cases[31,32,33]. The *RYrhoCentralFoam* solver and implementation of droplet submodels (e.g. droplet evaporation and multi-species diffusion) have been verified and validated with a series of benchmark cases against analytical solution and/or experimental data in our recent work[19,34].

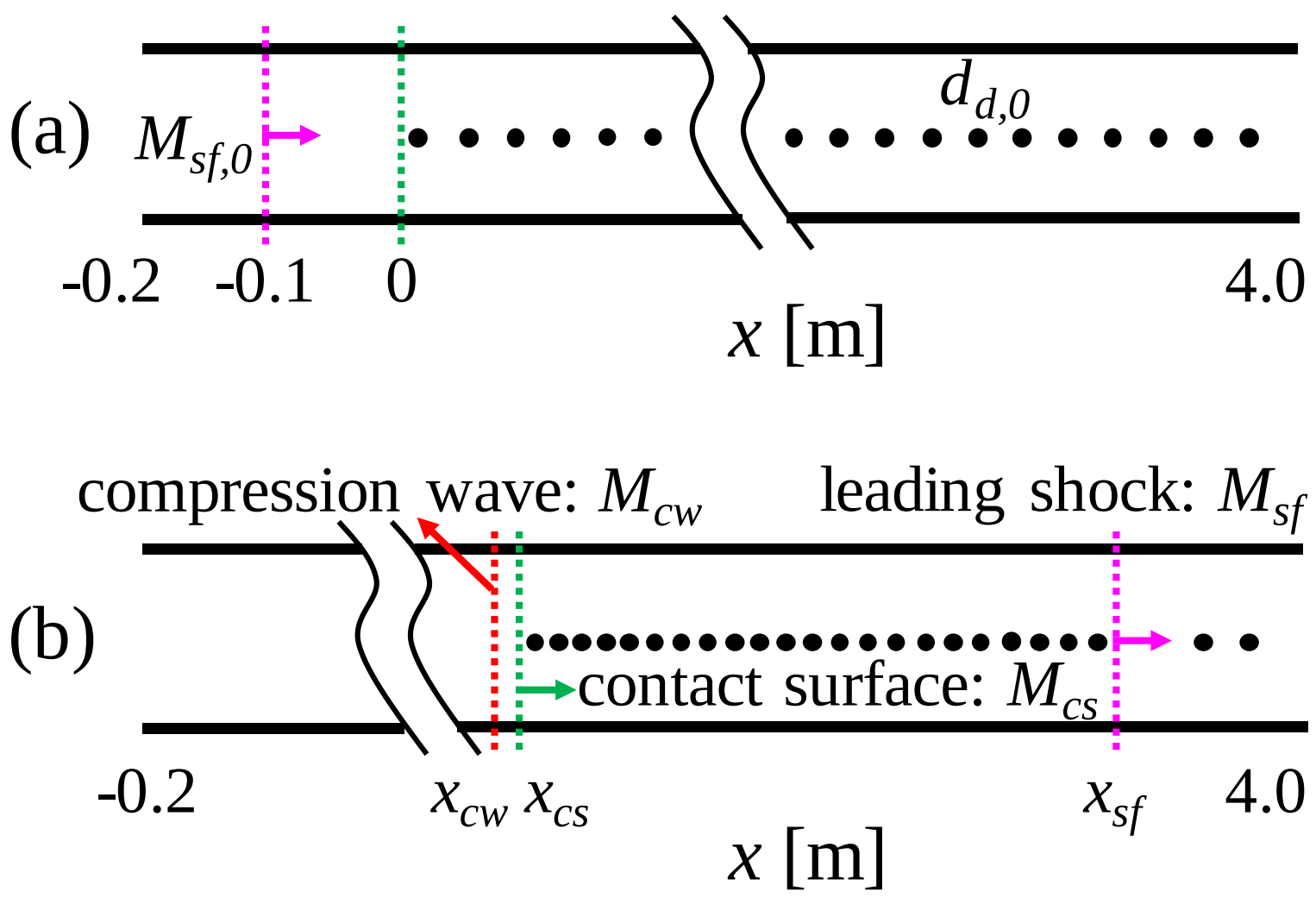

**Fig. 1.** Schematic of shock wave propagation in water droplet mists (a) before and (b) after shock enters the two-phase region. $M_{sf,0}$ is the Mach number of the incident shock initiated at $x$ = -0.1 m and $d_{d,0}$ is the initial diameter of water droplets. $M_{sf}$, $M_{cs}$ and $M_{cw}$ are the Mach numbers of leading shock, contact surface and compression wave. Circles: water droplets.

### 3. Physical problem

According to the experimental work of Hanson *et al.*[35], with spatially uniform aerosols, the flow behind the shock is nominally one-dimensional. Hence, one-dimensional shock wave propagation in water droplet mists is considered in this work. The one-dimensional simplification is also widely used for particle[36,37] or droplet[7,11,12,38] laden flows with shock waves. Figure 1(a) shows the schematic of the computational domain (which is 4.2 m in length and starts at $x$ = -0.2 m) and the initial distribution of the water droplets. The carrier gas is $O_2/N_2$ mixture with the mass fractions of 0.233 and 0.767, respectively. The right propagating shock is initiated at $x$ = -0.1 m with high pressure spot at $x$ < -0.1

m. In the pre-shock region ($x > -0.1$ m), the initial gas temperature and pressure are respectively 275 K and 66 kPa. These conditions are chosen consistently with the study of Goossens *et al.*[39] and Kersey *et al.*[7]. Four different incident shock Mach numbers are investigated, i.e. $M_{sf,0}$ = 1.17, 1.35, 1.5 and 1.6. The droplets are monodispersed and uniformly distributed in the region of $x > 0$ (i.e. the two-phase section). The considered initial diameters and number densities are 5−20 μm and $2.5 \times 10^{11} - 2 \times 10^{12}$ /m$^3$, respectively. The initial density, heat capacity and temperature of the water droplets are 1000.9 kg/m$^3$, 4222.4 J/kg/K and 275 K, respectively. Meanwhile, the water droplets are quiescent at $t = 0$. Figure 1(b) shows an instantaneous scenario after the shock propagates into the two-phase gas-droplet region. Three characteristic fronts are observable from our simulations, i.e. the leading shock, contact surface (interface of the purely gaseous and droplet-laden regions) and compression wave (interface of the expansion wave and shocked gas).

**Table 1.** Mesh sensitivity analysis.

| Cases # | M1 | M2 | M3 |
|---|---|---|---|
| $\Delta_L$ [mm] | 1.0 | 0.3 | 0.1 |
| $N_c$ | 4,200 | 14,000 | 42,000 |
| $N_{dc,0}$ | 10 | 3 | 1 |
| $N_{d,0}$ [/m$^3$] | | $5 \times 10^{11}$ | |

## 4. Mesh sensitivity analysis

The 1D domain in Fig. 1 is discretized with three meshes of 4,200, 14,000 and 42,000 cells. They respectively correspond to uniform cell sizes of 1.0, 0.3 and 0.1 mm, termed as meshes M1, M2 and M3. One droplet-laden case with shock Mach number $M_{sf,0}$ = 1.6 and initial droplet diameter $d_{d,0}$ = 5 μm is selected for mesh sensitivity analysis, which are detailed in Table 1. $N_c$ is the number of CFD cells, $\Delta_L$ is the uniform cell size, $N_{dc,0}$ is the initial droplet number in the cell. Note that although the initial numbers of droplet per CFD cell are different, the total number and the distributions of droplets in the entire domain are identical in the three cases, and the sensitivity of the statistical average in Eqs. (5)−(8) with respect to grid resolution will be studied.

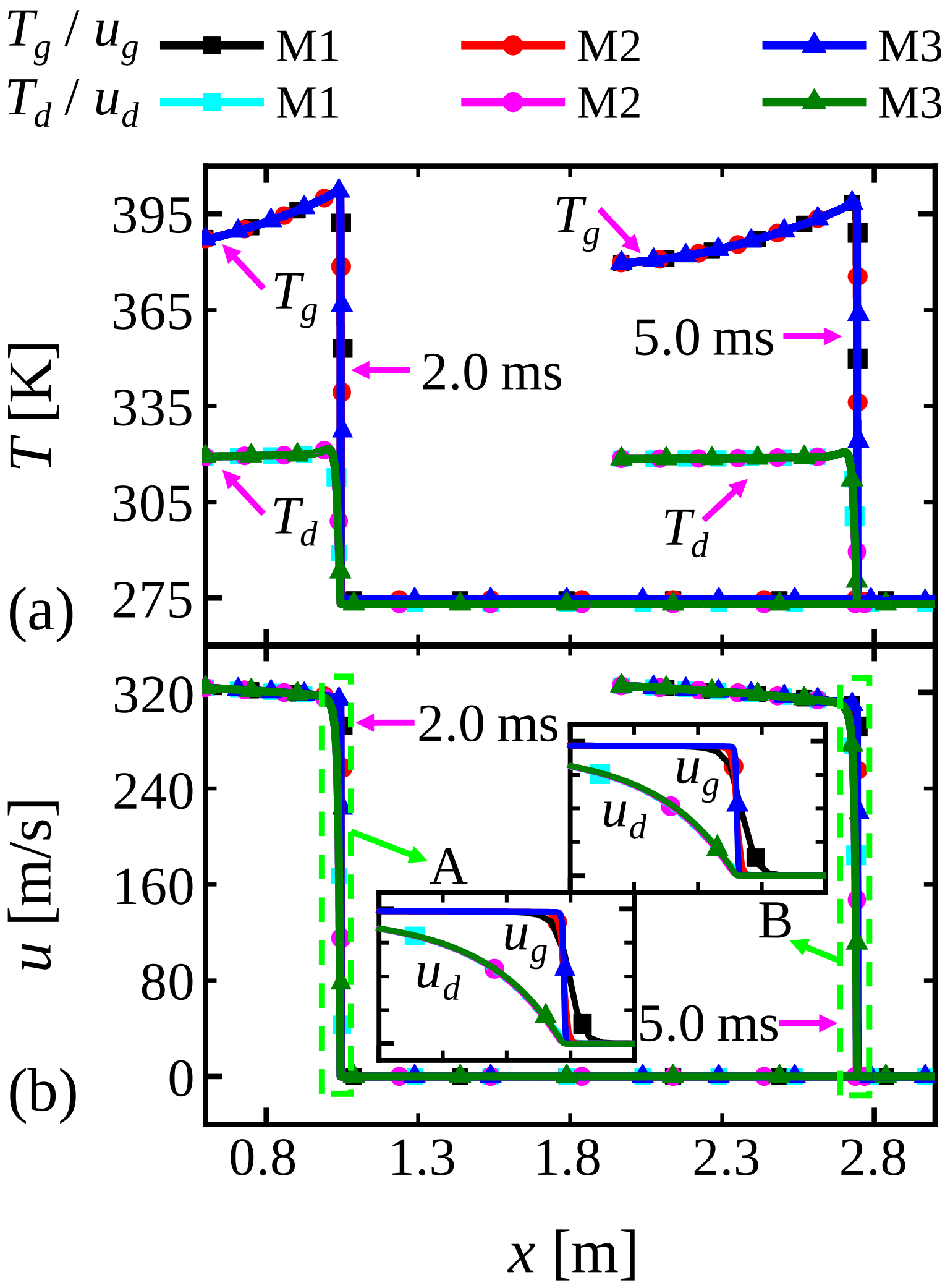

**Fig. 2.** Comparisons of (a) temperature and (b) velocity of gas and droplet phases with meshes M1, M2 and M3. $M_{sf,0}$ = 1.6, $d_{d,0}$ = 5 μm.

Figure 2 shows the profiles of temperature and velocity for both carrier gas and droplets at $t$ = 2.0

and 5.0 ms for Cases M1−M3. All the shown variables are from the Lagrangian results ($T_g$ and $u_g$ are interpolated to the droplet position). The near-shock regions for $t$ = 2.0 and 5.0 ms are enlarged in the insets of Fig. 2(b), which respectively range from $x$ = 1.03−1.05 m and $x$ = 2.73−2.75 m. It is seen that the differences in the temperature and velocity of two phases ($T_g$, $T_d$, $u_g$ and $u_d$) with different meshes are negligible. This is also true for other gas and droplet properties, e.g. gas viscosity, density, pressure and droplet diameters. Closer inspection of the profiles at $t$ = 2.0 and 5.0 ms reveals that all meshes accurately capture the shock front. Hence, both gas and droplet behaviors are not sensitive to the Eulerian mesh resolutions.

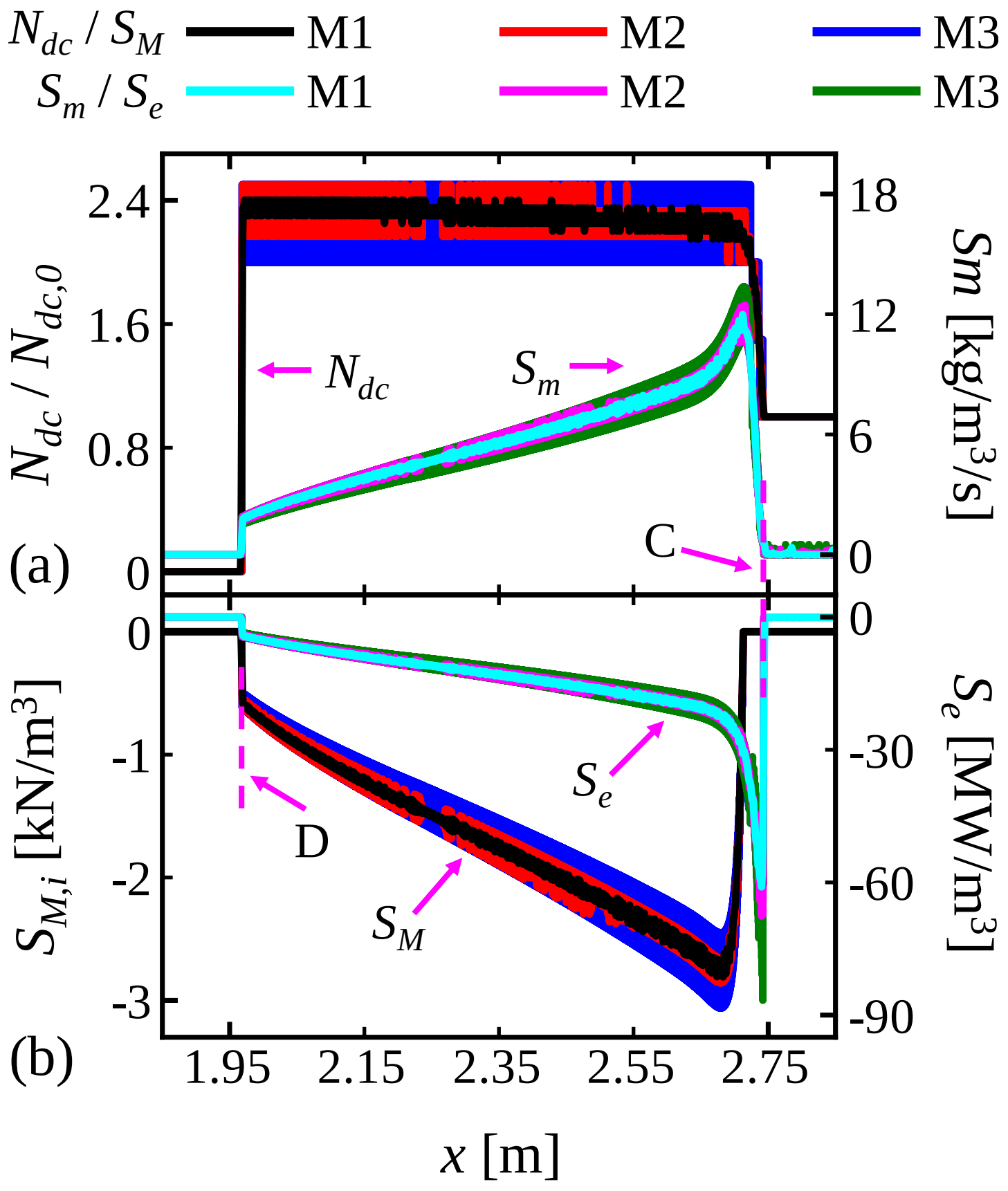

**Fig. 3.** Comparisons of (a) number of droplets per cell and mass exchange term, (b) momentum and energy exchange terms with meshes M1, M2 and M3. $M_{sf,0}$ = 1.6, $d_{d,0}$ = 5 μm.

Figure 3 shows the profiles of exchange terms in mass, momentum and energy equations (Eqs. 5−7) for Cases M1−M3. The numbers of droplets per cell (normalized by the corresponding $N_{dc,0}$) are also shown. For $S_m$, $S_{M,i}$ and $S_e$, the results from M3 have strong fluctuations in the regions between the leading shock and the contact surface, respectively indicated by the dashed lines C and D, although their averaged profiles are close with different meshes. Here, the average is based on the number of droplets per cell, i.e. divided by $N_{d,c}$. The fluctuations of $S_m$, $S_{M,i}$ and $S_e$ are caused by the variations of $N_{dc}$. For the finest mesh M3, droplet movement may cause significant change of the droplet number in a cell. This is confirmed by $N_{dc}$ distribution in Fig. 3(a), where the fluctuations in $N_{dc}$ increase with mesh resolution. These findings are also true for other shock Mach numbers. Based on Figs. 2 and 3, M1 will be used for the following analyses.

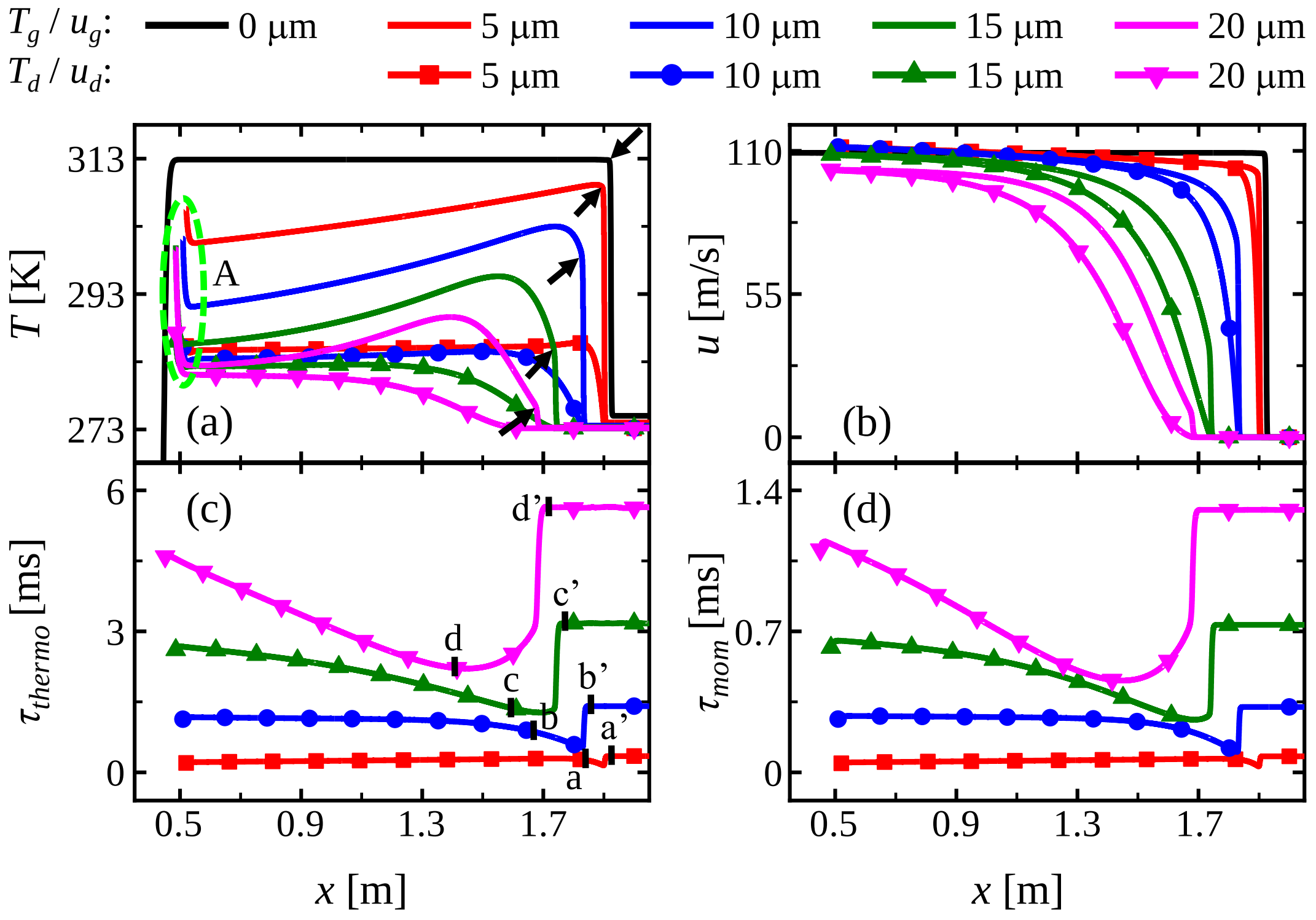

**Fig. 4.** Comparisons of (a) temperature and (b) velocity for gas and droplet phases, (c) droplet thermal and (d) momentum response timescales at $t$ = 5 ms for different droplet diameters.

## 5. Results and discussion

### *5.1. Effects of initial droplet diameter*

The effects of the initial water droplet diameter are studied in this section. Four diameters are considered, i.e. $d_{d,0}$ = 5, 10, 15 and 20 μm. The shock Mach number and droplet number density are fixed to be $M_{sf,0}$ = 1.17 and $N_{d,0}$ = 5 × $10^{11}$ /$m^3$, respectively. The resulting initial droplet volume fractions are $V_{fd,0}$ = 0.82 × $10^{-5}$, 6.55 × $10^{-5}$, 22.09 × $10^{-5}$ and 52.36 × $10^{-5}$, respectively. Figures 4(a) and 4(b) show the profiles of temperature and velocity for both gas and droplets at a representative instant, i.e. $t$ = 5 ms. The results from droplet-free case are also shown for comparison. It is seen that $T_g$ is more significantly decreased in the post-shock region (i.e. behind the arrows in Fig. 4a, which indicate the shock positions) with increased $d_{d,0}$. This is because larger droplets can absorb more heat from the gas phase, due to convective heat transfer and droplet evaporation. Note that the droplet equilibrium temperature near the two-phase contact surface are close for different cases, but it takes longer time to reach the equilibrium value for larger $d_{d,0}$. The gradient of $T_g$ near the leading shocks also gets weaker with increased droplet size and the original shock degrades to sonic wave when $d_{d,0}$ = 20 μm. Note that there is a local increase in $T_g$ in the locations marked by the ellipse A in Fig. 4(a). This corresponds to the travelling compression wave between the expansion wave and shocked gas, which is originated from $x$ = -0.1 m at $t$ = 0 s (see Fig. 1a). Therefore, higher (lower) total pressure and lower (higher) total temperature of this compressive wave can be observed, relative to its right (left) side. Similar increase in gas temperature near the contact surface also has been observed by Kersey *et al.*[7]. A more thorough examination of the wave structures in the flow field resulting from the interaction of a planar shock (two-dimensional) with a cloud of droplets has been performed by Chauvin *et al.*[8]. In their work, more complex wave structures and wave-droplet interactions are caused by the end-walls of the shock tube facility, e.g. extra expansion waves from the driver end-wall, reflected shock or compression waves from the driven end-wall. However, basic wave structures, including incident shock, compression wave and contact surface, are similar to ours.

It is seen from Fig. 4(b) that the shock is little affected by small droplets (e.g. 5 μm), due to the low volume fraction. For larger droplets (e.g. 20 μm), the equilibrium gas velocity $u_{g,eq}$ (the final velocity at the two-phase contact surface) even cannot be recovered to that of droplet-free case. The velocity equilibrium also takes longer distance with increased $d_{d,0}$. The gradient for gas velocity at the shock front are considerably reduced, indicating that the pronounced shock attenuation. This is confirmed by Fig. 5, which shows the evolutions of the instantaneous shock wave strength ($S_{sf}$, measured as the maximum magnitude of pressure gradient, $|\nabla p_g|_{max}$) and Mach number ($M_{sf}$, calculated as the propagation velocity of leading shock divided by the local speed of sound). It is seen from Fig. 5(a) that the strength of the propagating shock decreases in the droplet-laden gas, which is more remarkably for larger droplets. This can be confirmed by the consistently reduced shock Mach number. When $d_{d,0}$ = 20 μm, the leading shock is reduced to an acoustic wave with $M_{sf} \approx 1$ when $x > 2.3$ m.

Figures 4(c) and 4(d) respectively show the characteristic timescales for thermal and momentum response for all the droplets at $t$ = 5 ms. For a droplet in saturated gas, there is no evaporation and hence only convective heat transfer proceeds. Based on Eqs. (11), (26) and (27), the droplet temperature equation is reduced to

$$c_{p,d}\frac{dT_d}{dt} = \frac{6Nu \cdot k_g}{\rho_d \cdot d_d^2}\left(T_g - T_d\right). \tag{29}$$

Integration of Eq. (29) yields the thermal response time

$$\tau_{thermo} = \frac{c_{p,d} \cdot \rho_d \cdot d_d^2}{6Nu \cdot k_g}. \tag{30}$$

For a droplet with $Re_d << 1$, based on Eqs. (10) and (23)−(25), the momentum response time is expressed as

$$\frac{1}{\tau_{mom}} = \frac{18\mu_g}{\rho_d \cdot d_d^2} - \frac{1}{\rho_d}\left(\frac{\partial p}{\partial x} \cdot \frac{1}{u_g - u_d}\right). \tag{31}$$

In Figs. 4(c) and 4(d), for the unperturbed droplets in front of the leading shock (i.e. before points a'-d' in Fig. 4c), their characteristic timescales are proportional to the square of diameters as seen in Eqs. (30) and (31). Note that the pressure gradient is zero for these droplets. At the shock front, there is sharp decrease in both $\tau_{thermo}$ and $\tau_{mom}$, mainly due to the increased Nusselt number ($Re_d$ increases sharply, see Eq. 27) and the pronounced PGF at the shock front, respectively. After then (e.g. after points a-d in Fig. 4c), $\tau_{thermo}$ and $\tau_{mom}$ increase and gradually levels off with increased off-shock distance. Hence, the effects of leading shock on the characteristic timescales only lie in a narrow region (e.g. a-a' for $d_{d,0}$ = 5 μm in Fig. 4c), and the width of this region decreases with $d_{d,0}$. In some work[36,37,40,41], such region is termed as "relaxation zone", in which significant momentum and energy exchanges occur between the two phases until the equilibrium is reached. Furthermore, $\tau_{thermo}$ is generally 4-6 times larger than $\tau_{mom}$ for the same droplet in these cases, indicating that, for dispersed droplets, accommodation of the velocity to that of the continuous phase is generally much faster than that of temperature. This can be further confirmed in Figs. 7 and 9 later.

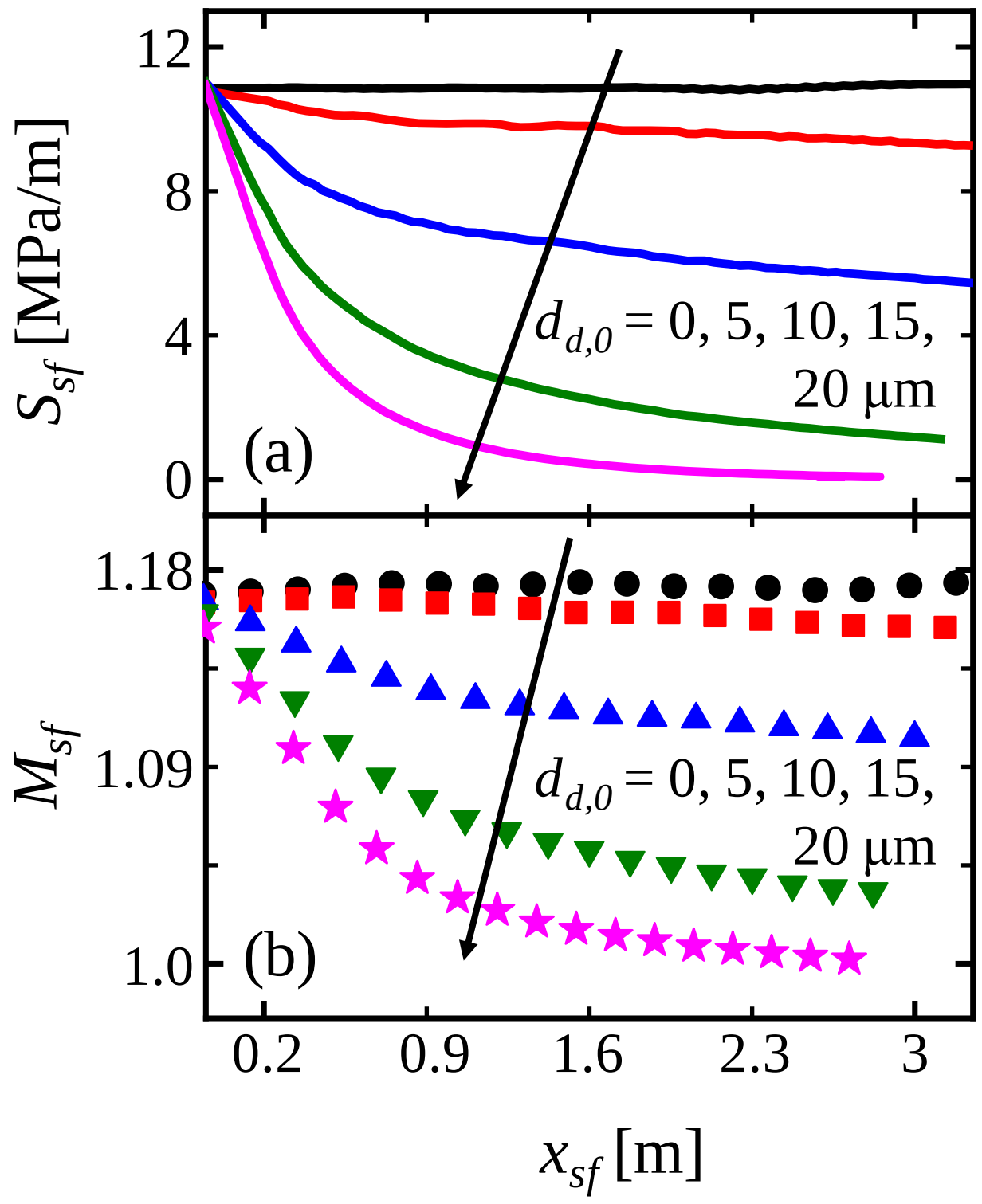

**Fig. 5.** Evolutions of shock (a) strength and (b) Mach number.

Figure 6 shows the evolutions of droplet volume fraction in $x-t$ diagram with four initial droplet diameters (5, 10, 15 and 20 μm). At the right side of lines A−D, the droplets are intact and hence the volume fractions are spatially uniform. Meanwhile, at the left side of lines A'−D', it is droplet-free and the volume fraction is zero. It is interesting to find from Fig. 6(a) that the droplets of $d_{d,0}$ = 5 μm can be quickly accelerated to the local gas speed due to their small momentum response times (see Fig. 4d and Eq. 31, the first term on RHS of Eq. 31 is large for small $d_{d,0}$), which hence makes the droplet volume fraction high immediately behind the leading shock. For $d_{d,0}$ = 10 μm in Fig. 6(b), the droplets that freshly enter the shocked region has not responded slowly to the local gas speed, due to the increased momentum response time (see Fig. 4d). This leads to a transition distance with unchanged droplet volume behind the leading shock. However, further downstream, the volume fraction is almost uniform, which means that the droplets are propagating at the close velocities. For larger droplets ($d_{d,0}$ = 15 μm), the above transition distance is wider. This is more obvious when $d_{d,0}$ = 20 μm, in which most of the droplets are accumulated at the contact surface (line D' in Fig. 6d).

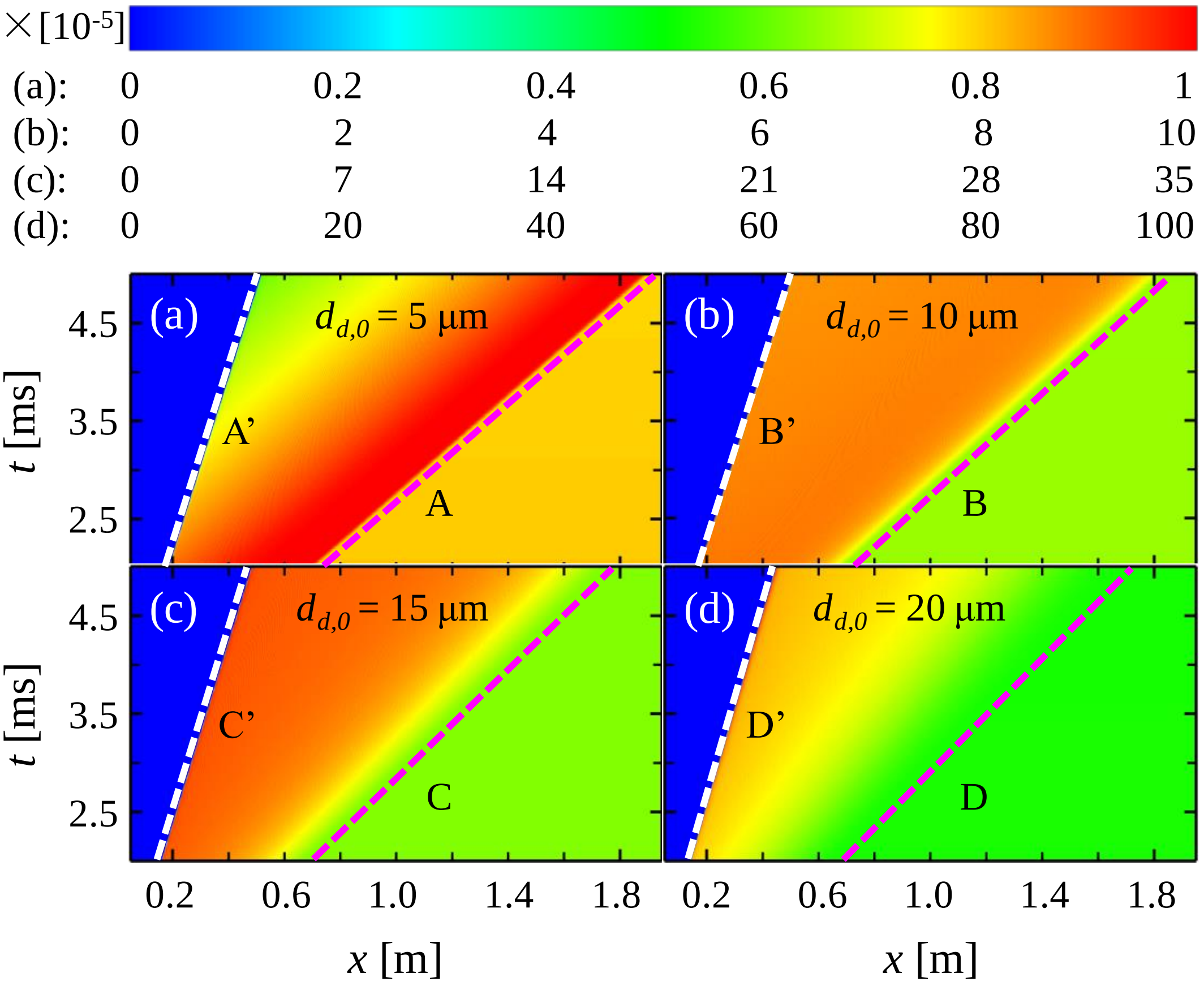

**Fig. 6.** $x$-$t$ diagram of droplet volume fraction. A−D: leading shock; A'−D': contact surface.

Figure 7 shows the evolutions of the momentum exchange term ($S_{M,i}$ in Eq. 6) for the gas phase in $x-t$ diagram for the above four cases. It is seen that interphase momentum exchange is completed immediately behind the leading shock when the initial droplet diameter is 5 μm. Note that $S_{M,i} \approx 0$ in the droplet-laden region means almost no interphase velocity difference and hence all the droplets share the same speed to the gas. Also, the momentum relaxation zone is very narrow and the width is almost constant with respect to time. However, this zone is extended for larger droplets (e.g. 10, 15 and 20 μm). This leads to almost uniform distribution of droplet volume fraction after some distance of the leading shock in Fig. 6 especially for $d_{d,0}$ = 5 and 10 μm. In Figs. 7(c) and 7(d), the momentum relaxation becomes more distributed. One feature, different from that with small-sized droplets, is that the momentum exchange (particularly the intensity, as shown in Figs. 7c and 7d) varies with time. This is caused by the evolving response of the droplets to the gradually weakened shock waves in these two cases as indicated in Fig. 5.

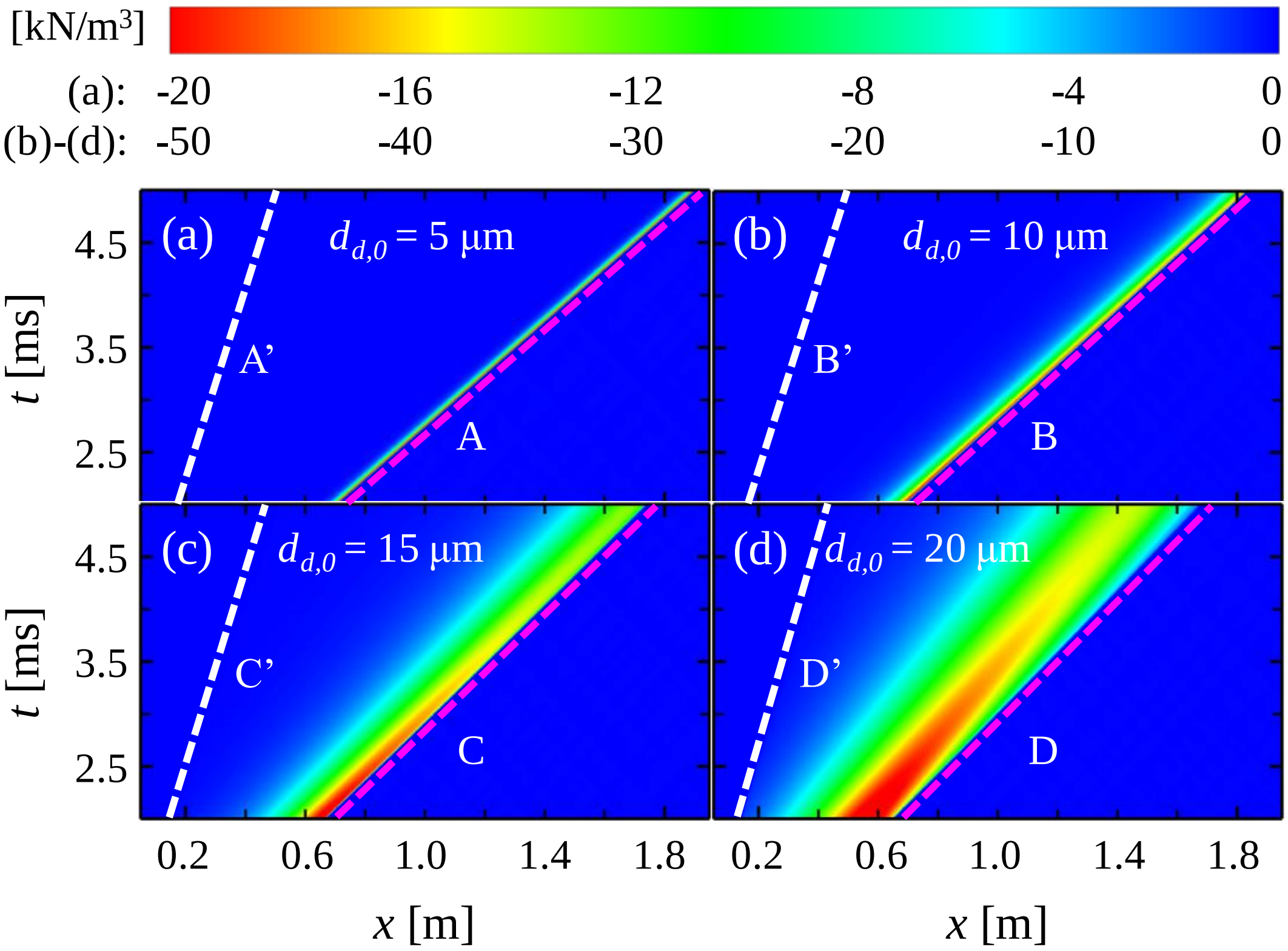

**Fig. 7.** *x-t* diagram of momentum exchange term with droplet diameters of (a) 5, (b) 10, (c) 15 and (d) 20 μm. Legend for dashed lines same as in Fig. 6.

Figure 8 shows the evolutions of mass exchange term (i.e. volumetric evaporation rate, $S_m$ in Eq. 5) in *x-t* diagram when $d_{d,0}$ = 5, 10, 15 and 20 μm. For $d_{d,0}$ = 5 μm, strong evaporation occurs closely behind the leading shock and decreases gradually in the post-shock region. This is caused by the slow decrease in temperature after the leading shock (see Fig. 4a) and small interphase velocity difference (see Fig. 7a), which would result in moderate interphase heat transfer (shown in Fig. 9a later) for droplet evaporation. In the work of Watanabe *et al.*[42], the front where evaporation first occurs behind the leading shock is termed as evaporation front. It is also observable in our cases as significant evaporation generally lags behind the leading shock in Fig. 8. For $d_{d,0}$ = 10 and 15 μm, droplet evaporation becomes strong after a finite distance behind the leading shock. The distributions of the evaporation zones are almost constant with respect to time. For $d_{d,0}$ = 20 μm, within the shown period, droplet evaporation is considerably reduced, due to much lower gas temperature (see Fig. 4). The carrier gas cannot provide sufficient energy for droplet evaporation when the droplet volume fraction is sufficiently high.

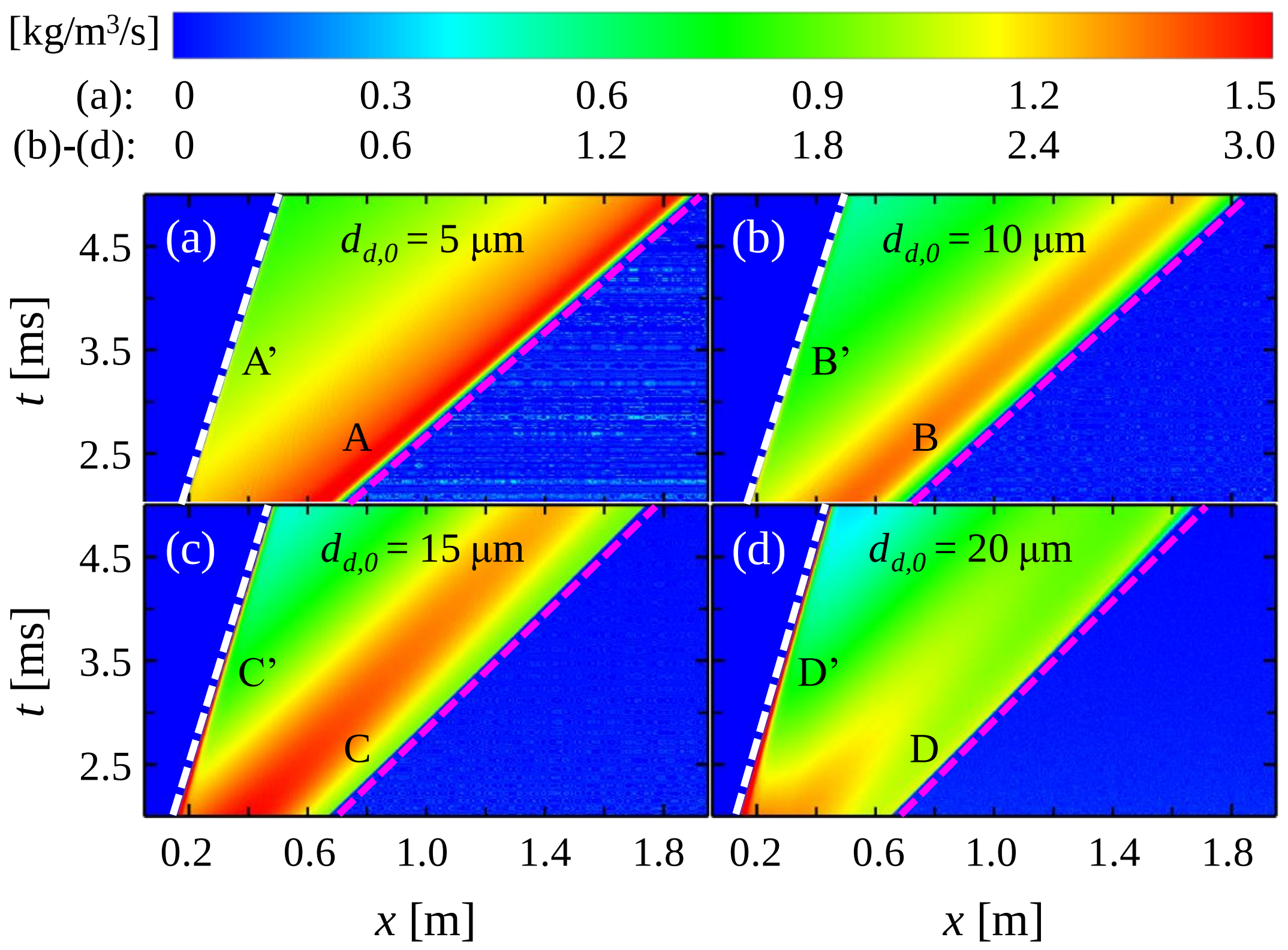

**Fig. 8.** *x-t* diagram of mass exchange term with droplet diameters of (a) 5, (b) 10, (c) 15 and (d) 20 μm. Legend for dashed lines same as in Fig. 6.

Figure 9 shows the evolutions of energy exchange term ($S_e$ in Eq. 7) in $x-t$ diagram. Generally, considerable energy transfer occurs earlier than the mass exchange when the shock wave sweeps. This is because droplet heating towards the saturated temperature occurs due to convective heat transfer, which takes a finitely long period. When the droplets reach the saturated temperature, significant evaporation can be observed. Therefore, with the same droplet size, the energy exchange zones with high $|S_e|$ (e.g. above 8 MW/m$^3$ for $d_{d,0}$ = 5 μm and 16 MW/m$^3$ for $d_{d,0}$ = 10, 15 and 20 μm) are closer to the leading shock front than droplet evaporation zone. Meanwhile, the energy exchange zone is remarkably narrower than mass exchange zone for the droplets of $d_{d,0}$ = 5, 10 and 15 μm. This implies that once the droplets are heated to high temperatures (relative to their initial temperature), they are self-sustainable for continuously evaporate with large $S_m$ but with low $|S_e|$.

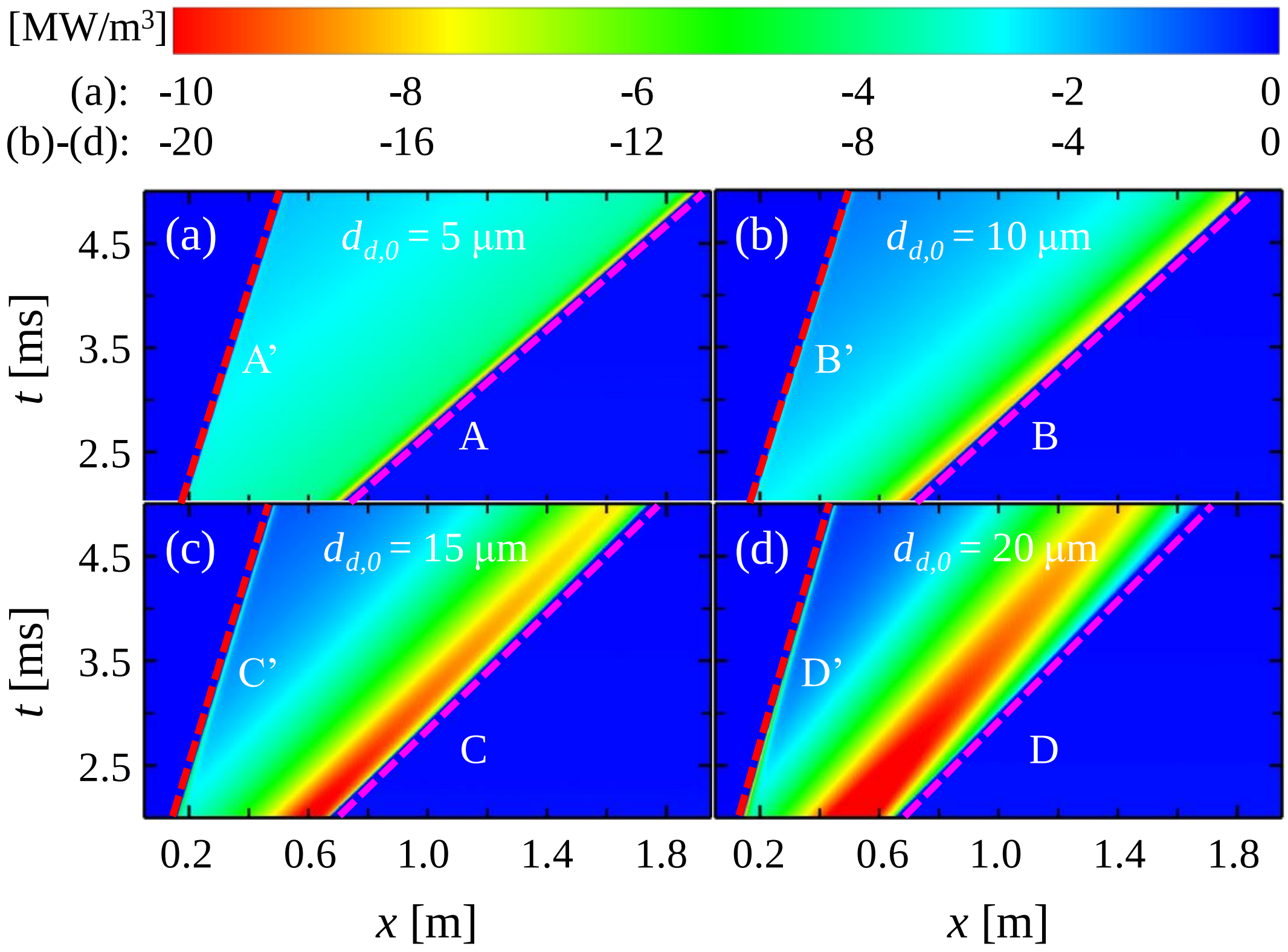

**Fig. 9.** *x*-*t* diagram of energy exchange term with droplet diameters of (a) 5, (b) 10, (c) 15 and (d) 20 μm. Legend for dashed lines same as in Fig. 6.

The spatial and temporal differences of interphase mass, momentum and energy exchanges is worthy of further discussion. For fine water droplets (e.g. 5 μm), the timings for these exchanges are close. For large water droplets (e.g. $d_{d,0}$ = 10 and 15 μm), the difference between onsets of energy and momentum exchanges, as well as mass exchange, is more pronounced. This is because $S_{M,i}$ mainly depends on the motions of gas and droplets. Note that the contribution to $S_{M,i}$ from the first term (the momentum transfer from the evaporated vapor) on the RHS of Eq. (6) is small compared to the other two terms, as the volumetric evaporation rate $S_m$ is generally four orders of magnitude smaller than that of $|S_{M,i}|$ (see Figs. 7 and 8). However, variations of $S_m$ and $S_e$ depend on different local conditions, e.g. vapor saturability, interphase temperature difference and droplet heating. This indicates that when the leading shock interacts with the droplets, the interphase momentum exchange is the major reason of shock attenuation, as considerable mass and energy exchanges lag remarkably farther behind the shock front. Hence, the shock attenuation is mainly affected by the instantaneous amount of droplets it interacts with.

*5.2. Effects of droplet number density*

The effects of droplet number density on the propagating shock wave will be examined in this section. Figure 10 shows the evolutions of gas phase total pressure ($p_{g,tot}$) in $x-t$ diagram with four different droplet number densities, i.e. $N_{d,0}$ = 2.5 × $10^{11}$, 5 × $10^{11}$, 1 × $10^{12}$ and 2 × $10^{12}$ /m$^3$. Their corresponding droplet volume fractions are 3.27 × $10^{-5}$, 6.55 × $10^{-5}$, 13.09 × $10^{-5}$ and 26.18 × $10^{-5}$, respectively. Here the shock wave Mach number and initial droplet diameter are $M_{sf,0}$ = 1.35 and $d_{d,0}$ = 10 μm, respectively. In Figs. 10(a)-10(d), the total pressure in the regions between lines A' and A", B'-B", C'-C" and D'-D" (here A'-D' denote the contact surfaces, A"-D" denote the compression waves) is continuous to that behind lines A"-D". This means that the contact surface, not the compression wave, is a jump interface for gas phase total pressure. Also, $p_{g,tot}$ monotonically increases in the droplet-laden areas, from A to A'. This is caused by the total pressure recovery from droplet evaporation, which acts as mass addition to the gas phase. Moreover, the larger droplet number density $N_{d,0}$ leads to stronger recovery, as can be seen from Figs. 10(a)-10(d). This can be further confirmed by the contours of $Y_{H2O}$ in Fig. 12. However, the total pressure $p_{g,tot}$ is relatively low immediately after the leading shock (lines A−D). This is caused by two factors, i.e. the shock-induced and droplet-induced (due to both drag and pressure gradient forces) gas pressure loss. Specifically, the initially quiescent droplets act as strong aerodynamic barriers to the incident shock. Furthermore, the droplet-induced pressure loss becomes stronger with increased $N_{d,0}$. This is because that both drag and pressure gradient forces from droplets increase with droplet numbers that immediately interact with the shock front (see Eqs. 23 and 25, respectively). Note that although pressure can be recovered to some extent due to droplet evaporation, however, it cannot reach the original value, i.e. $p_{g,tot}$ in the droplet-laden area is always lower than that of the driver gas (left to lines A''-B'').

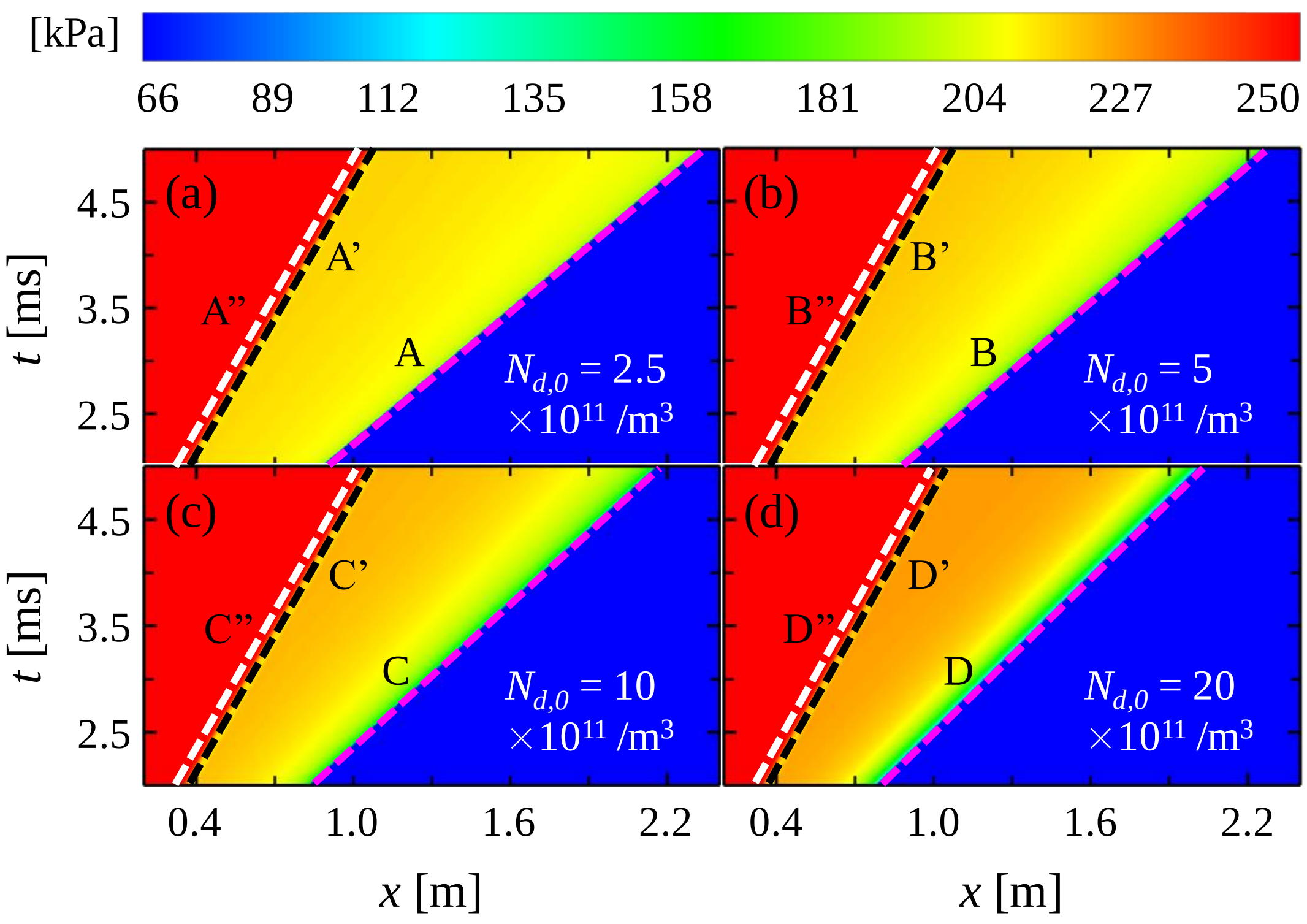

**Fig. 10.** *x*-*t* diagram of gas total pressure with various droplet number densities. A-D: leading shock; A'-D': contact surface; A''-D'': compression wave.

Figure 11 shows the evolutions of gas phase total temperature ($T_{g,tot}$) in $x-t$ diagram for the same cases as in Fig. 10. When the droplet number density is $N_{d,0}$ = 2.5 × $10^{11}$ /m$^3$, $T_{g,tot}$ varies slightly behind the leading shock (almost constant around 363 K). This is because relatively weak heat exchange between the two phases. As $N_{d,0}$ further increases, $T_{g,tot}$ is considerably reduced due to heat transfer to the dispersed droplets. Meanwhile, with larger $N_{d,0}$, more significant $T_{g,tot}$ reduction is observable at the end of the post-shock evaporation zone. This would slow down droplet evaporation and interphase heat transfer. Furthermore, $T_{g,tot}$ also decreases due to shock attenuation ($Ma_g$ ↓, not shown here as already demonstrated in Fig. 5 and will be further confirmed in Fig. 14), droplet heat absorption ($T_g$ ↓) and vapor mass addition ($\gamma_g$ ↓) based on that $T_{g,tot} = T_g\left(1 + \frac{\gamma_g - 1}{2} Ma_g^2\right)$, with $Ma_g$ and $\gamma_g$ being the Mach number and specific heat ratio of the carrier gas, respectively.

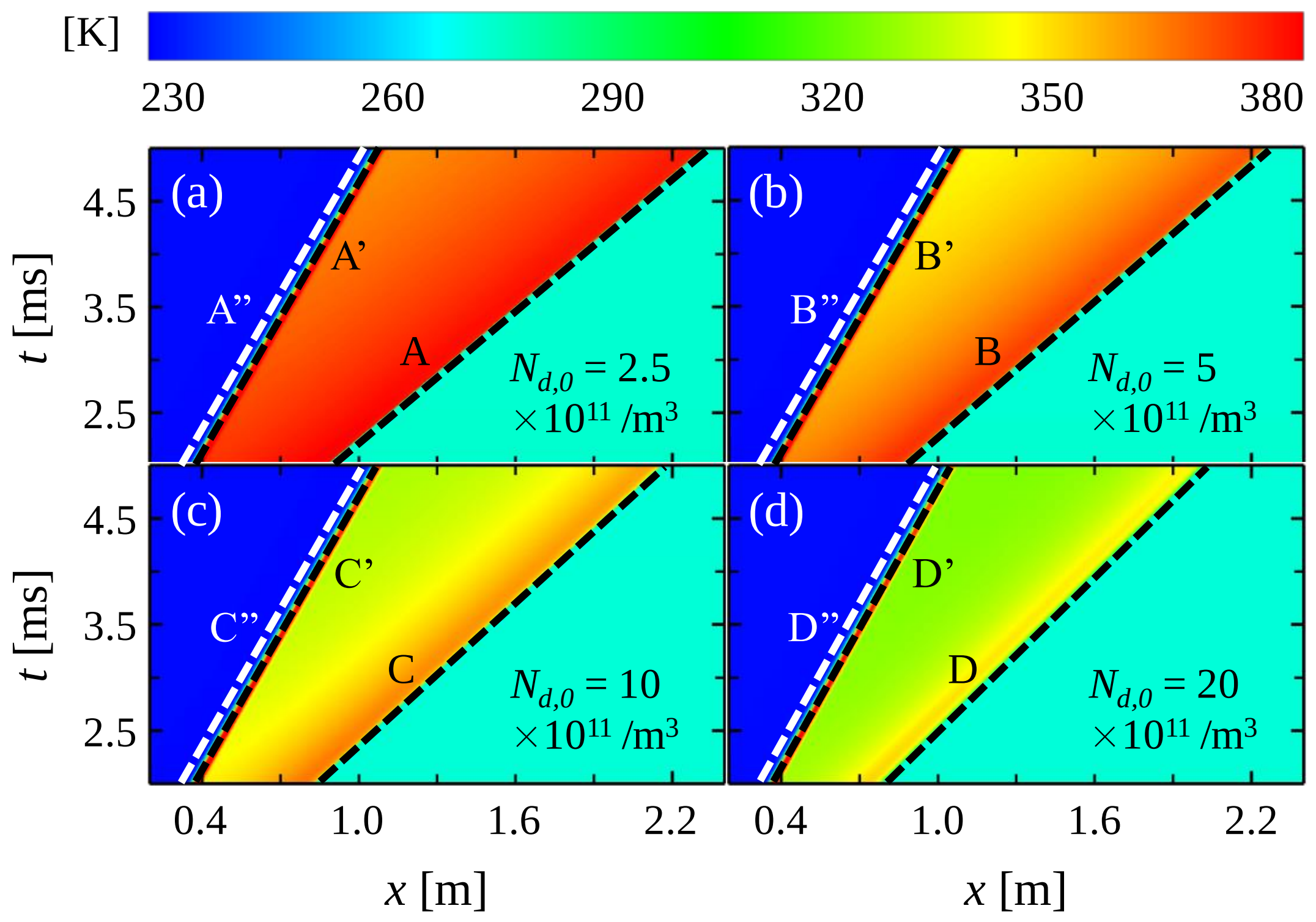

**Fig. 11.** *x-t* diagram of gas total temperature with various droplet number densities. Legend for dashed lines same as in Fig. 10.

Figure 12 shows the evolutions of $H_2O$ mass fraction in $x-t$ diagram. Generally, $Y_{H2O}$ increases towards the contact surface, which is accumulated from continuous evaporation. For $N_{d,0} = 2.5 \times 10^{11}$ and $5 \times 10^{11}$ /m$^3$, the volumetric evaporation rate $S_m$ (see Fig. 13a, at $t$ = 5.0 ms for example) is relatively low, whereas the gas is unsaturated even at $t$ = 5.0 ms (see Fig. 13b). However, for $N_{d,0}$ = 1 $\times 10^{12}$ and $2 \times 10^{12}$ /m$^3$, considerable vaporization (hence $S_m$) is found in the shocked gas and water vapor concentration near the contact surface is high, as seen in Fig. 13. This is particularly true when $N_{d,0} = 2 \times 10^{12}$ /m$^3$ in Fig. 12(d), in which most of the area between D and D' is filled with $H_2O$ vapor of high concentration. This can be confirmed by the shorter distance between where remarkable water concentration exists and the leading shock front (indicated by lines A−D in Fig. 12).

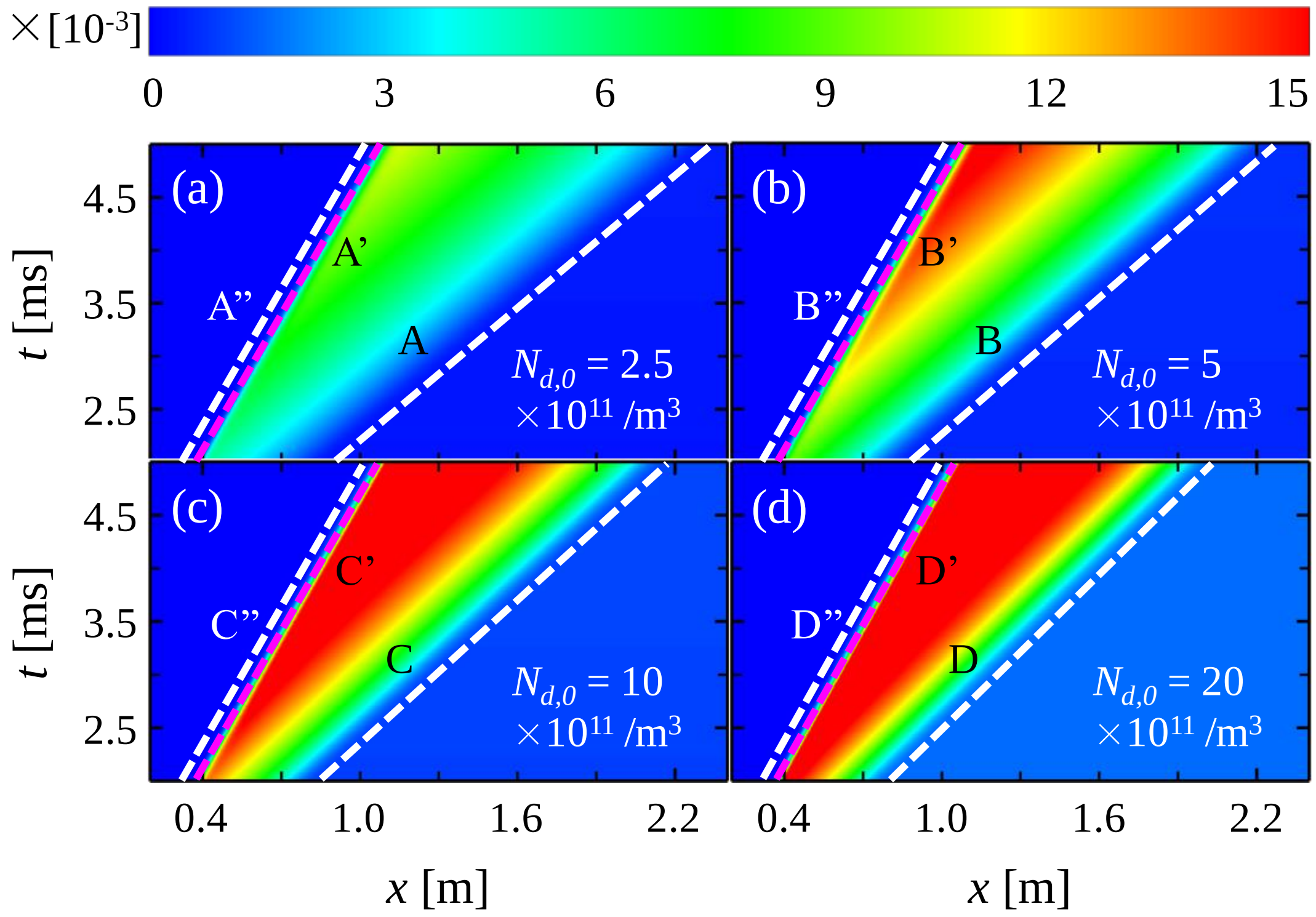

**Fig. 12.** *x-t* diagram of $H_2O$ mass fraction with various droplet number densities. Legend for dashed lines same as in Fig. 10.

It is observed from Figs. 10-12 that the compression wave (A"-D") is always on the left side of the contact surface (A’-D’) that separates the two-phase region from the pure gas region, and therefore the droplets cannot cross the compression waves. This is due to the following reasons. Firstly, all the droplets have been accelerated to the local gas speed near the left end of the droplet-laden region (i.e. lines A’-D’ in the *x-t* diagrams above). Hence, there is no slip velocity between the compression waves and the local droplets. Secondly, the compression wave results in a strong positive pressure gradient, which would act on and therefore accelerate the droplets if the droplets approaches the compression wave.

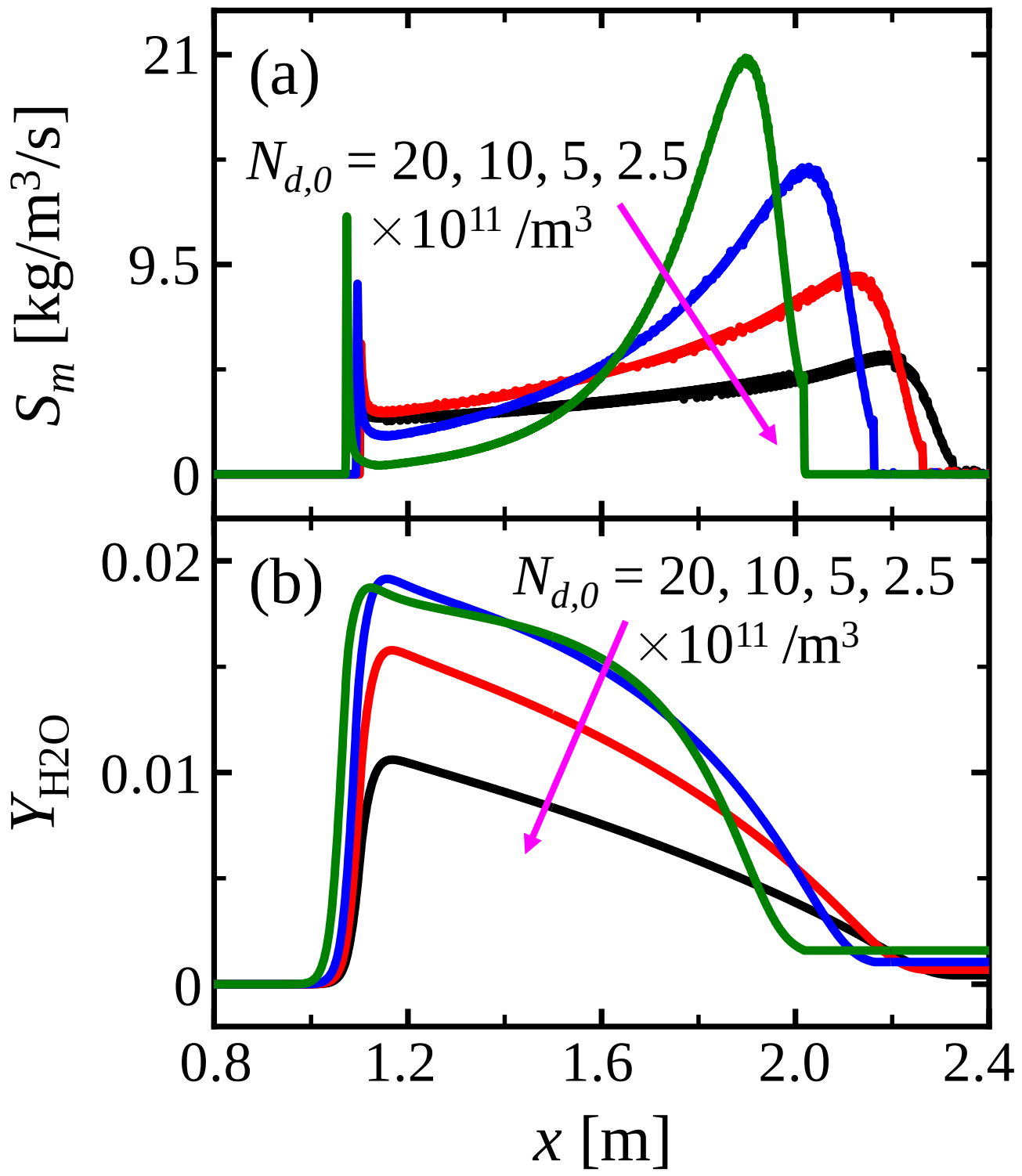

**Fig. 13.** Comparisons of (a) mass exchange term and (b) $H_2O$ mass fraction at $t$ = 5 ms for different droplet number densities.

The shock attenuation by the dispersed droplets has been shown in Fig. 5. For a specific droplet volume fraction, the shock velocity is attenuated to the same value at the same location when it propagates downstream, regardless of the droplet size[38]. Figure 14 further studies this effect, through visualizing the shock Mach numbers at different streamwise locations, subject to various initial droplet volume fractions. For $V_{fd,0} < 10^{-4}$, the shocks are slightly attenuated when they propagate, as $M_{sf}$ at different locations are quite close. Obvious shock attenuation occurs when $V_{fd,0} > 10^{-4}$. It is found from Fig. 14(a) that, with initial Mach number $M_{sf,0} = 1.17$, $M_{sf}$ quickly decays to less than 1.0 for $x > 1.0$ m when $V_{fd,0} > 4 \times 10^{-3}$. For higher $V_{fd,0}$ (e.g. $V_{fd,0} > 8 \times 10^{-3}$), $M_{sf}$ is lower than 1.0 even before $x = 0.5$ m. For $M_{sf,0} = 1.35$ in Fig. 14(b), $M_{sf} > 1$ for $x = 0$-$3.0$ m when $V_{fd,0} < 4 \times 10^{-3}$. With further increased $V_{fd,0}$, the shocks finally evolve to pressure waves. Similar observation has been reported by Kersey *et al.*[7], in which an incident shock with $M_{sf,0} = 1.25$ is attenuated to subsonic when the droplet mass loading is above 0.63. From Fig. 14, it is also seen that the shocks with higher $M_{sf,0}$ can propagate farther downstream with larger $V_{fd,0}$ before they are attenuated to sonic waves.

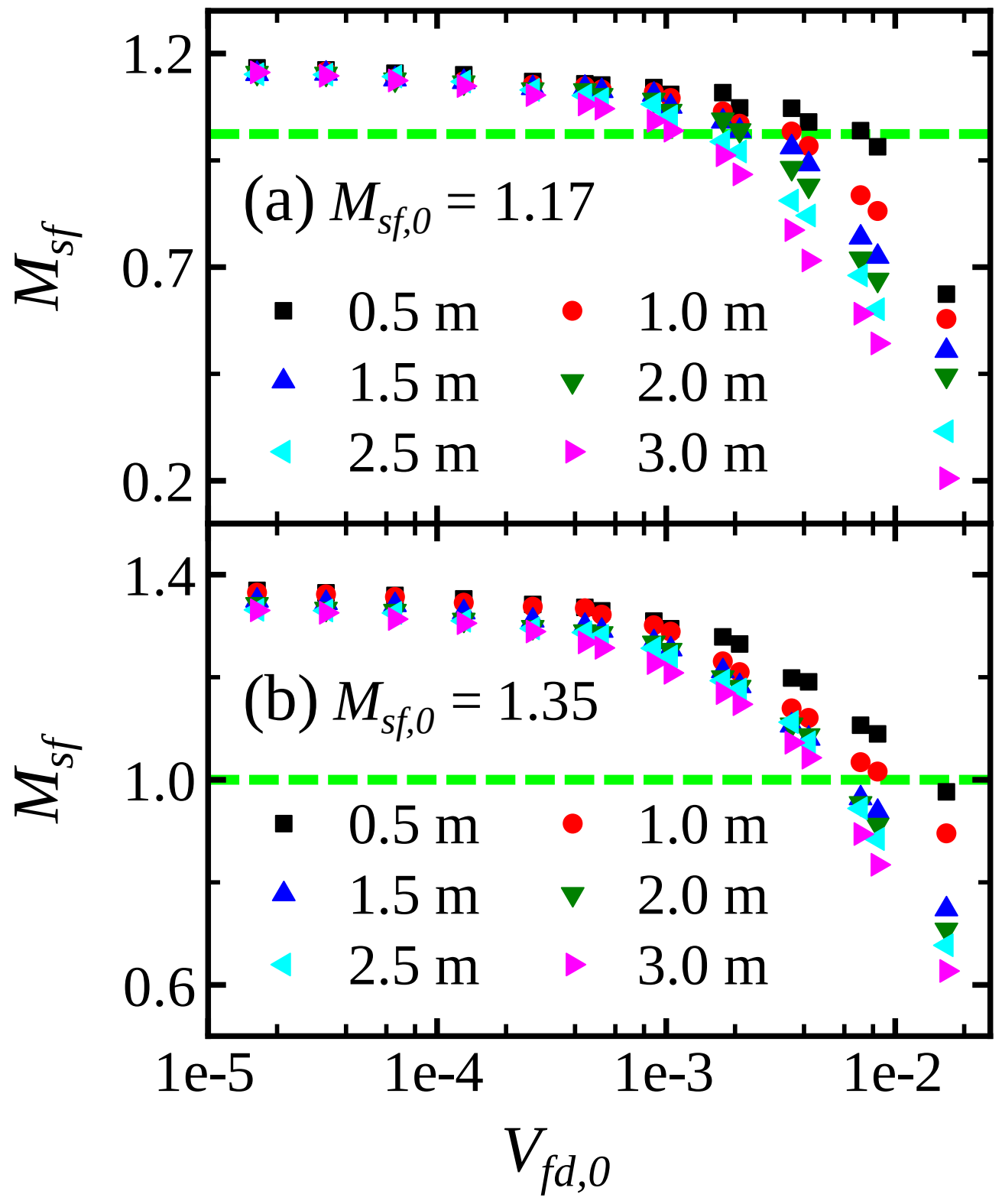

**Fig. 14.** Shock Mach numbers with various initial droplet volume fractions: (a) $M_{sf,0} = 1.17$ and (b) $M_{sf,0} = 1.35$. Dashed lines: iso-lines of $M_{sf} = 1.0$.

*5.3. Effects of incident shock Mach number*

Figure 15 shows the evolutions of droplet diameters in $x-t$ diagram with four incident shock Mach numbers, i.e. $M_{sf,0}$ = 1.17, 1.35, 1.5 and 1.6. The corresponding compression wave Mach numbers are $M_{cw,0}$ = 0.31, 0.61, 1.2 and 1.5. Apparently, the compression wave is supersonic when $M_{sf,0}$ = 1.5 and 1.6. The initial droplet diameter, number density and volume fraction are $d_{d,0}$ = 5 μm, $N_{d,0} = 1 \times 10^{12}$ /m$^3$ and $V_{fd,0} = 1.64 \times 10^{-5}$, respectively. Generally, in the shocked gas the droplet diameter decreases continuously from the evaporation front. Near the leading shock (lines A-D), the droplet diameter is about $d_d$ = 4.95 μm, which is slightly smaller than $d_{d,0}$ = 5 μm due to the slow evaporation in the quiescent air before the shock arrival. It is seen that $d_d$ decreases faster with larger $M_{sf,0}$ due to the increased interphase temperature difference. The smallest droplets are observed at the end of the two-phase region (lines A'-D'), due to their relatively larger residence time in the shocked gas and hence longer evaporation time. Note that the two-phase contact surfaces when $M_{sf,0}$ = 1.5 and 1.6, i.e. C' and D', are demarcated with a turning point (i.e. e and f) into two sections (see Figs. 15c and 15d). The probable reason for this phenomenon is that the contact surface (line C' and D') and compression wave (lines C" and D") behind the leading shock are supersonic in these two cases, and will be explained in detail below.

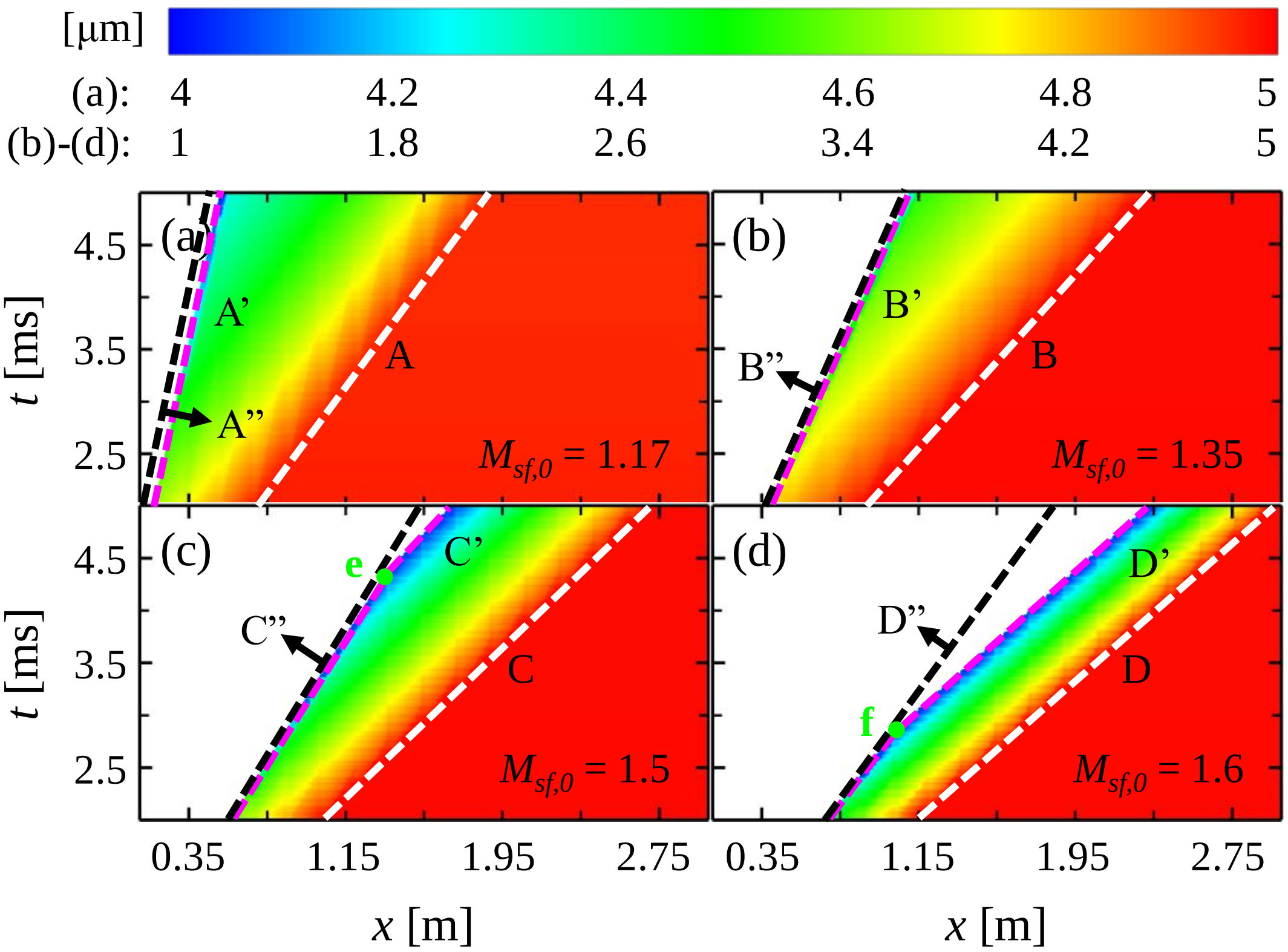

**Fig. 15.** *x*-*t* diagram of droplet diameters with various incident Mach numbers. Legend for dashed lines same as in Fig. 10.

The recompression phenomenon can be justified from the distributions of static pressure $p_g$ and Mach number $Ma_g$, which are shown in Fig. 16. The results are from a representative instant $t$ = 5 ms. In Fig. 16(a) with $M_{sf,0}$ = 1.5 and 1.6, it is seen that pressure behind the expansion wave (points a'' and b'' in Fig. 16b) is higher than that after the shock (points a and b in Fig. 16b) and compression (points a' and b' in Fig. 16b) waves. The higher pressure behind the expansion wave accelerates the compression wave to catch up with the leading shock, which is decelerated by the droplets, till the droplets and shock wave reach a kinematic balance. After that, the leading shock and contact surface propagate at the same speed. This is why the leading shock is parallel to the contact surface in Figs. 15(c) and 15(d), after point e in $M_{sf,0}$ = 1.5 and point f in $M_{sf,0}$ = 1.6, respectively. However, this phenomenon is not seen with lower incident Mach number (e.g. $M_{sf,0}$ = 1.17 and 1.35). In Fig. 16(b), it is seen that the gas after the compression wave (but before the expansion wave) is supersonic in $M_{sf,0}$ = 1.5 and 1.6, but subsonic in $M_{sf,0}$ = 1.17 and 1.35. Note that recompression of the leading shock is only possible when the compression wave is supersonic. Although there is also acceleration of the compression wave to some extend (e.g. to 0.36 Ma and 0.81 Ma at $t$ = 5 ms, respectively) when $M_{sf,0}$ = 1.17 and 1.35, they do not reach supersonic propagation, and hence has little effects on the leading shocks. Furthermore, the smaller difference between $M_{sf,0}$ and $M_{cw,0}$ is, the faster $M_{sf}$ attenuates to $M_{cw}$ (therefore the earlier they balance), as seen in Fig. 15.

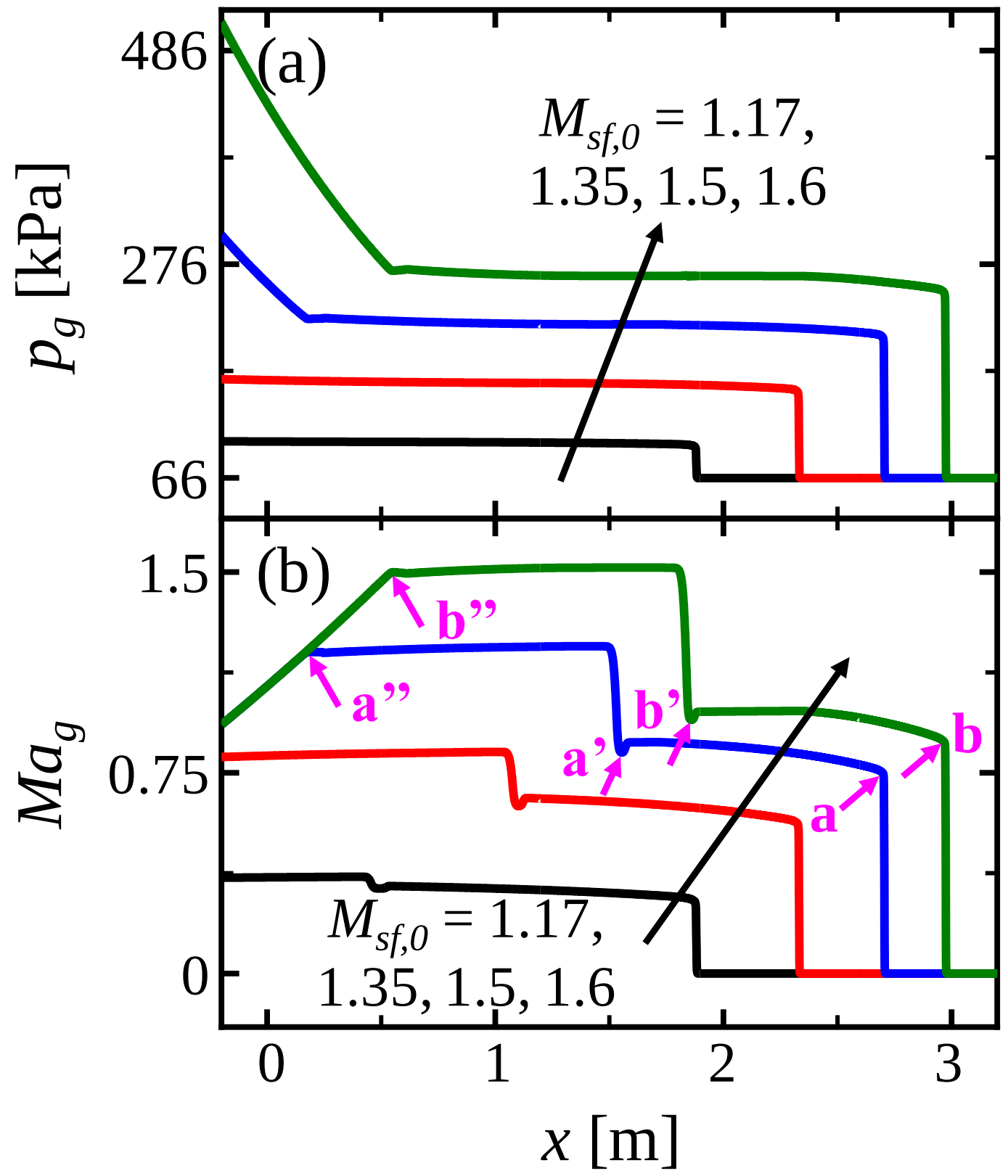

**Fig. 16.** Distributions of (a) pressure and (b) Mach number at $t$ = 5.0 ms for different incident shock Mach numbers. a/b: shock front; a’/b’: compression wave; a”/b”: expansion wave.

The interphase coupling behind the peculiar recompression process is further discussed in Figs. 17(a)-17(c), which show the evolutions of exchange terms for mass, momentum and energy equations for the gas phase from the same cases in Figs. 15 and 16. The results are volume-integrated in the droplet-laden region. For $M_{sf,0}$ = 1.17 and 1.35, $S_m$, $S_{M,i}$ and $S_e$ change monotonically. However, for $M_{sf,0}$ = 1.5 and 1.6, before their turning points (i.e. e and f in Fig. 17) $S_m$, $S_{M,i}$ and $S_e$ vary in similar tendency to those with lower shock Mach numbers. However, after the turning points, variations of $S_m$, $S_{M,i}$ and $S_e$ are limited. It is seen that $S_m$ and $|S_e|$ decrease slowly and almost linearly with time after the turning points, which indicates slightly weakened evaporation and heat transfer. This is because the region between the leading shock and contact surface (i.e. the region between lines C and C', D and D' in Fig. 15) approaches saturation (see the constant $H_2O$ vapor mass fraction of about 0.0193 at $t$ = 5 ms, circled by ellipses G and H in Fig. 17d) after the turning points. This is particularly true in the region close to the contact surface due to the higher shocked gas temperature, which develops faster to approach saturation condition. Hence, the main contributions to $S_m$ and $S_e$ come from the droplets that just enter the shocked region. However, as the water droplets continuously enter, the gas velocity and temperature of this region slightly decrease with time, which then leads to slow decrease in $S_m$ and $|S_e|$. Also, obvious momentum exchange only comes from the droplets that freshly enter the shocked region. The droplets well behind the shock have reached the kinematic equilibrium with the ambient gas. Hence, $|S_{M,i}|$ is quite small as only the droplets close to the shock front have to be accelerated, and $|S_{M,i}|$ almost decreases linearly and weakly due to the insignificant shock attenuation. From Figs. 15 and 17, it is found that for these two cases, the region between the leading shock and contact surface reaches steady state after the turning point, not only in terms of its structural width, but also two-phase interactions, e.g. mass, momentum and energy exchange.

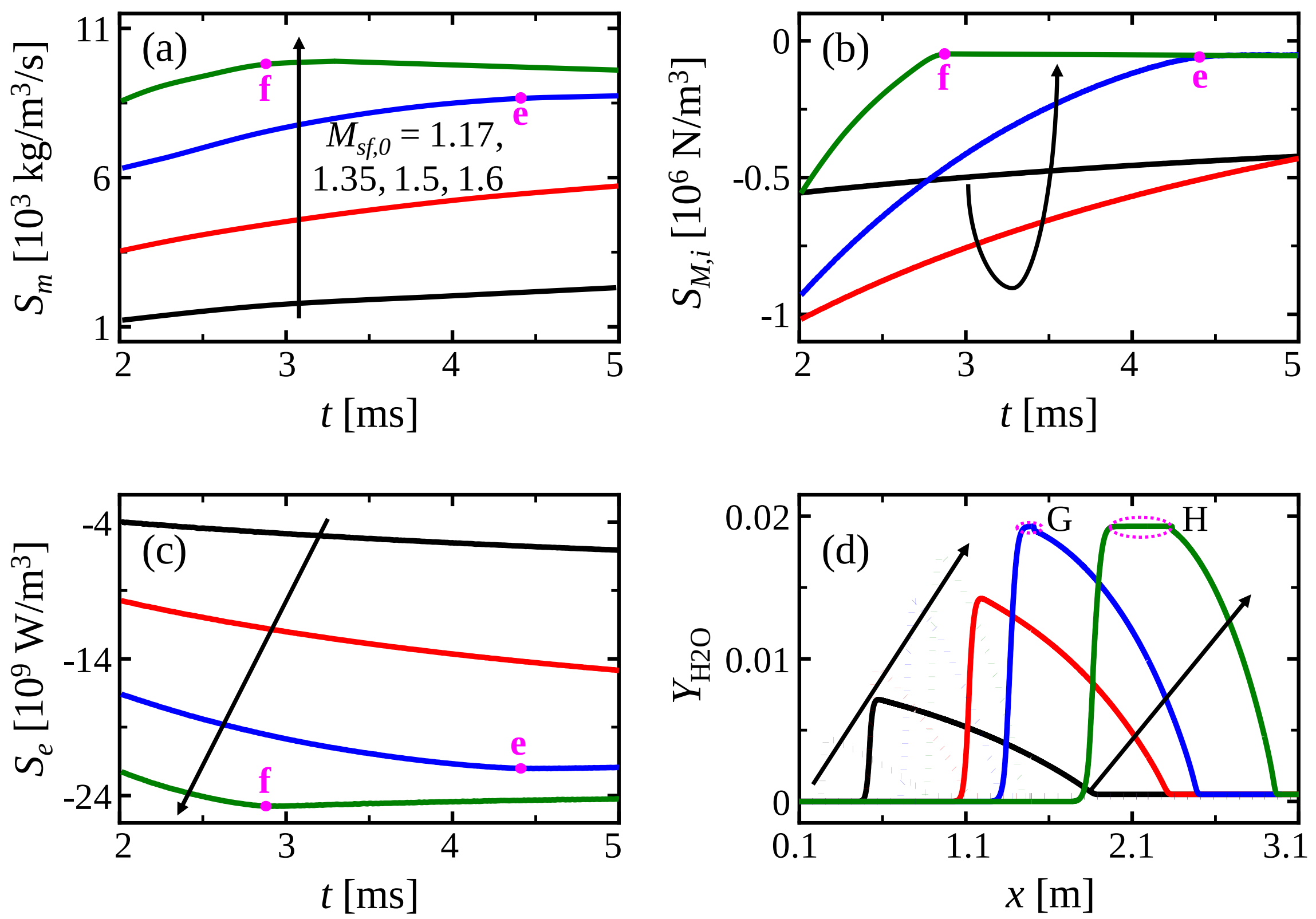

**Fig. 17.** Evolutions of volume-integrated (a) mass, (b) momentum and (c) energy exchange rates in the droplet-laden region with various incident shock Mach numbers. (d) Profiles of $H_2O$ mass fraction at $t$ = 2.5 ms (dot lines) and 5.0 ms (solid lines). Points e and f are the turning points in Fig. 15, ellipses G and H are vapor-saturated regions.

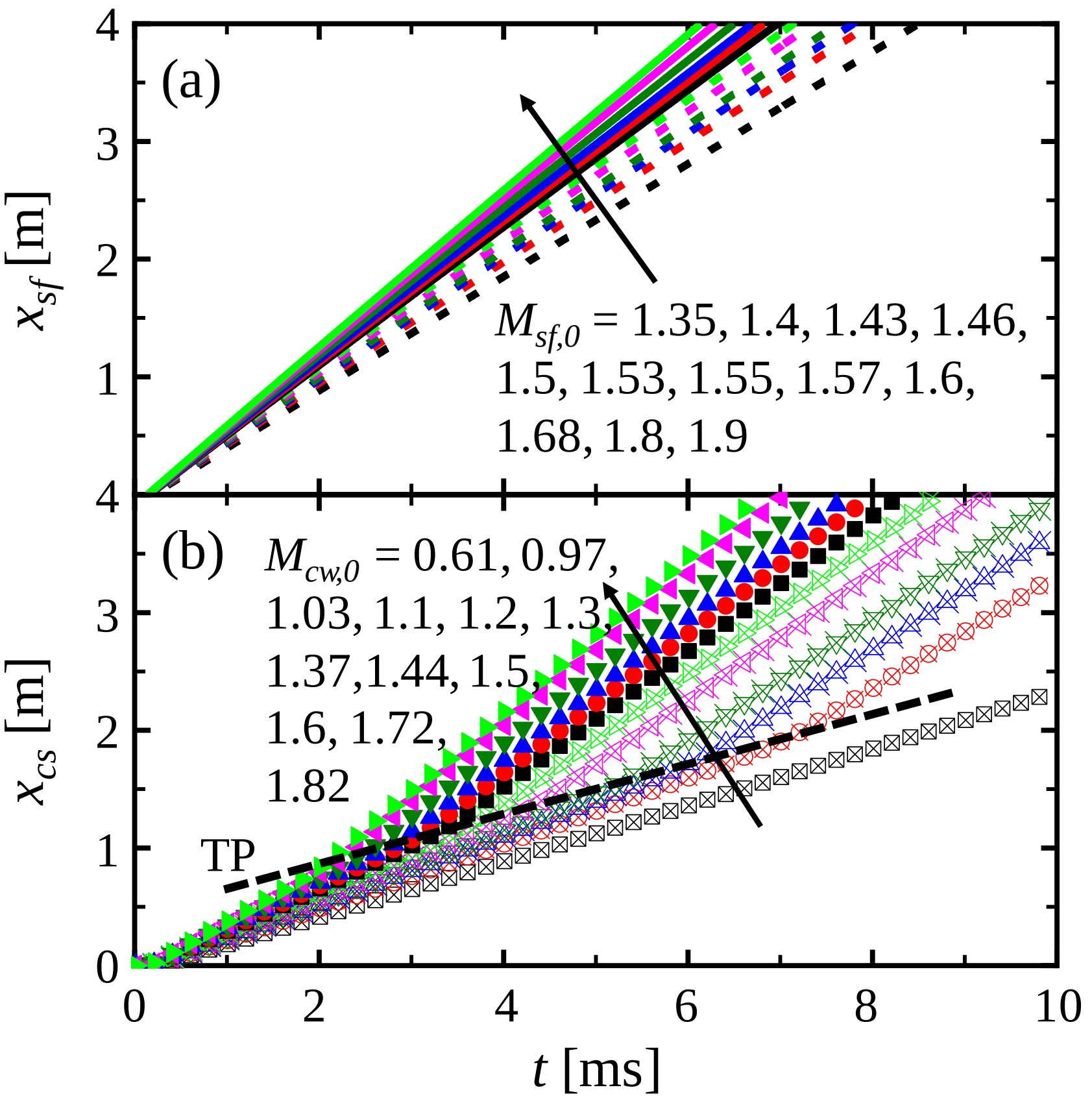

**Fig. 18.** Evolutions of leading shock and contact surface locations under various initial shock Mach numbers. The dashed line TP indicates the track of the turning points in different cases.

Based on the above analyses, it is found that whether the compression wave is supersonic is an indication of the recompression phenomenon. Figure 18 further shows the evolutions of the instantaneous locations of leading shock ($x_{sf}$) and contact surface ($x_{cs}$) under various initial shock Mach numbers. All the cases share identical initial droplet diameter and number density in Figs. 15-17, i.e. $d_{d,0}$ = 5 μm and $N_{d,0} = 1 \times 10^{12}$ /m$^3$. The initial leading shock and the corresponding compression wave Mach numbers are indicated in Figs. 18(a) and 18(b), respectively. It is seen that all cases with $M_{cw,0}$ > 1.0 can reach a balance between the leading shock and two-phase contact surface, i.e. their velocities (slopes of their profiles) are almost the same after the turning points. Furthermore, the larger $M_{sf,0}$ is, the earlier the turning point occurs, and the closer the contact surface to the leading shock, which has been confirmed in Figs. 15 and 17. Note that the case with $M_{sf,0}$ = 1.4 is different, the initial compression wave is subsonic ($M_{cw,0}$ = 0.97), however, the compression wave can be accelerated to be supersonic when the shock propagates. This acceleration has been shown in Fig. 16 for cases $M_{sf,0}$ = 1.17 and 1.35, mainly caused by the higher pressure in the left expansion wave than that in the compression wave before they reach an equilibrium (before the turning point). For cases with $M_{sf,0} \leq$ 1.35, $M_{cw}$ is subsonic throughout the computational domain although there is also acceleration of the compression wave (e.g. $M_{cw}$ from 0.61 initially to 0.81 at $t$ = 5 ms for $M_{sf,0}$ = 1.35 as seen in Fig. 16), but there is no recompression phenomenon.

Finally, Fig. 19(a) shows evolutions of the distance between leading wave and two-phase contact surface under different droplet volume fractions. The incident shock Mach number is fixed to be 1.6. In Fig. 19(a), for $V_{fd,0} = 0.82 \times 10^{-5}$ (the same case shown in Figs. 15-18 with $M_{sf,0}$ = 1.6) and $1.64 \times 10^{-5}$, recompression is observable after their respective turning points (f and g in Fig. 19a). For $V_{fd,0} = 3.28 \times 10^{-5}$, recompression may occur after $t \approx 5.5$ ms, as circled by ellipse I in Fig. 19(a). However, this weak recompression phenomenon is evolving (i.e. $x_{sf} - x_{cs}$ does not reach constant) even when the leading shock exits from the domain. For $V_{fd,0} \geq 13.12 \times 10^{-5}$, there is no recompression. Furthermore, the onset of the recompression phenomenon occurs later with larger initial droplet volume fraction (no recompression may be regarded as the turning point at $t \to +\infty$). Therefore, it can be seen that initial droplet volume fraction also affects the occurrence of the recompression phenomenon.

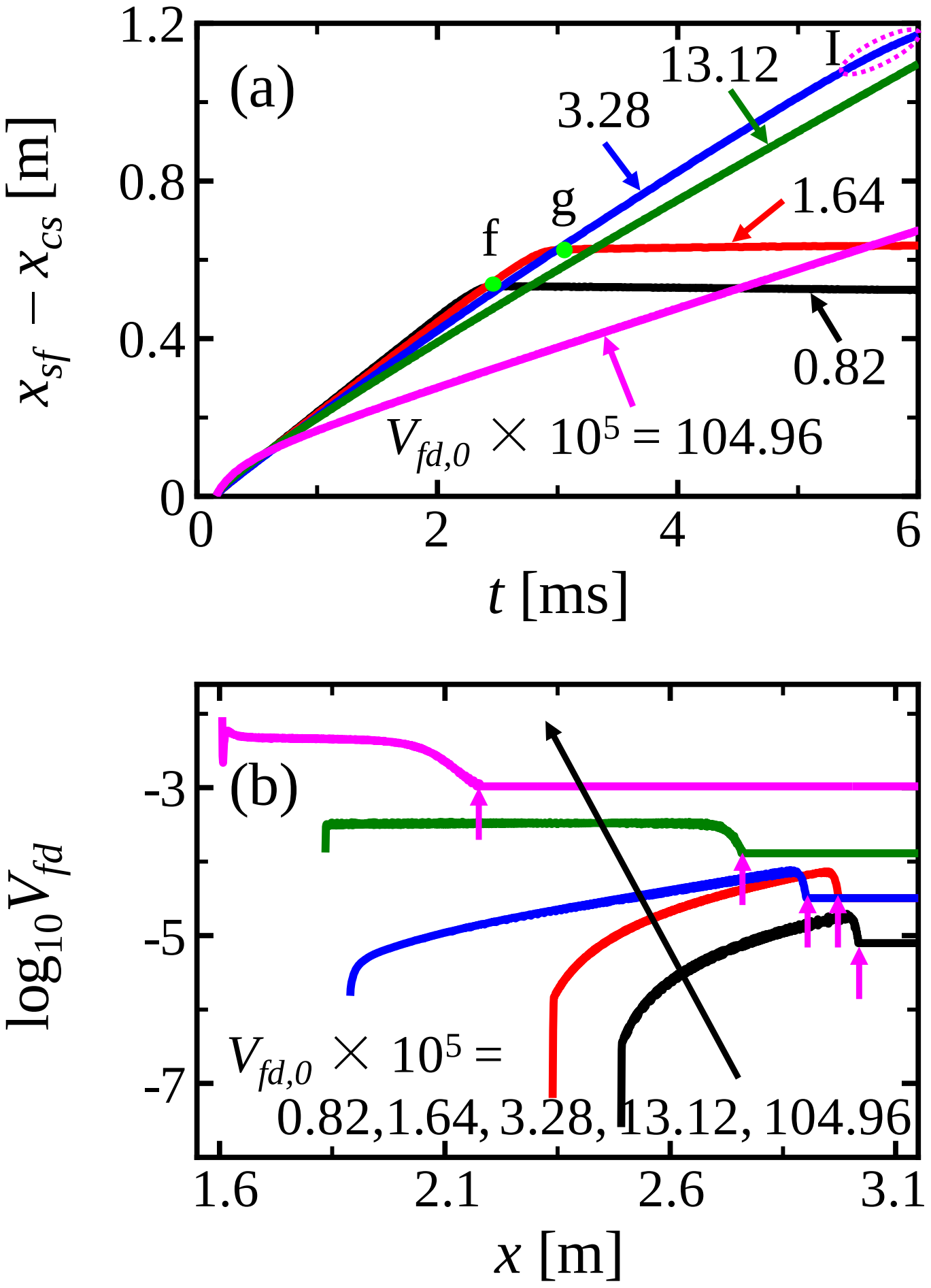

**Fig. 19.** Initial droplet volume fractions: (a) evolutions of the distance between leading shock wave and contact surface, (b) droplet volume fraction (in logarithmic scale) at $t = 5.0$ ms. The arrows in Fig. 19(b) indicate the instantaneous leading waves.

Figure 19(b) shows the corresponding droplet volume fraction at $t = 5.0$ ms. For $V_{fd,0} = 0.82 \times 10^{-5}$ and $1.64 \times 10^{-5}$, $V_{fd}$ is low (e.g. $\leq 10^{-7}$) at the contact surface (i.e. the left end of each profile). Furthermore, $V_{fd}$ decreases rapidly from its peak value after the leading shock (indicated by the arrow in Fig. 19b) to the contact surface. Therefore, attenuation of the supersonic compression wave behind the contact surface is weak based on Fig. 14 and recompression can occur. For $V_{fd,0} = 3.28 \times 10^{-5}$, it is seen that $V_{fd} > 10^{-6}$ at the contact surface, and this condition is critical for occurrence of the recompression process. For $V_{fd,0} \geq 13.12 \times 10^{-5}$, $V_{fd}$ gets higher at the contact surface and recompression has no chance to occur. Based on Figs. 18 and 19, one can see that the recompression phenomenon is sensitive to droplet volume fraction. For low $V_{fd,0}$ = 0.82 × $10^{-5}$, recompression can occur even with lower Mach number, e.g. $M_{sf,0}$ = 1.4 and $M_{sf,0}$ = 0.97. However, for higher droplet volume fraction $V_{fd,0}$ = 3.28 × $10^{-5}$, it is difficult to occur even at $M_{sf,0}$ = 1.6 and $M_{cw,0}$ = 1.5. Therefore, the condition for stable recompression is that the compression wave can be accelerated to supersonic condition by the followed high-pressure expansion wave (see Fig. 16), but not attenuated excessively with high droplet loading.

## 6. Conclusion

The interactions between propagating shock waves and dilute evaporating water droplets are investigated numerically in a one-dimensional domain. The exchanges in mass, momentum, energy and vapor species between the carrier gas and droplets are considered through two-way coupling of Eulerian-Lagrangian approach. Emphasis is laid on shock attenuation, two-phase interactions, droplet evaporation, motion and heating dynamics. Parametric study is performed for initial droplet diameters of 5 - 20 μm, initial droplet number densities of 2.5 × $10^{11}$ - 2 × $10^{12}$ /$m^3$, and incident shock Mach numbers of 1.17 - 1.9. Three characteristic fronts are observed when the shock travels in the two-phase gas-droplet medium: the leading shock that propagates right with fastest velocity, the contact surface that separates the droplet-laden region from droplet-free region, and the compression wave that follows behind the contact surface.

The variation of shock Mach numbers with droplet volume fractions indicates that remarkable shock attenuation generally takes place when the initial droplet volume fraction is larger than $10^{-4}$. Furthermore, the propagating shock may be attenuated to sonic or even subsonic pressure waves when the initial droplet volume fractions are larger than $10^{-3}$ and 6 × $10^{-3}$ for incident shock Mach numbers of 1.17 and 1.35, respectively. The evolutions of interphase exchanges in mass, momentum and energy indicate that mass and energy exchange is generally 4-6 times slower than momentum exchange for water droplets. Therefore, attenuation of the shock in both strength and propagation speed is mainly caused by the momentum loss to those droplets that interact with the shock front. Shock compression, drag force and pressure gradient force lead to remarkable total pressure loss immediately after the shock. However, in the shocked region with significant droplet evaporation, total pressure recovery is observed due to vapor addition into the carrier gas.

Gas recompression in the region between the leading shock and contact surface is found for high incident shock Mach numbers, which is because the attenuated leading shock reaches a balance with the following supersonic compression wave. When the recompression phenomenon occurs, the width of this recompressed region tends to be constant as the velocities of the leading shock (attenuated due to interactions with droplets) and compression wave (accelerated due to the residual of the high initial pressure for initiating the incident shock) are the same. Furthermore, the total amount of interphase exchanges in mass, momentum and energy in this region also tends to be steady. The turning point in the *x*-*t* diagram, after which the recompression phenomenon stabilizes, occurs earlier with increased incident shock Mach number. For an initial droplet volume fraction of $1.64 \times 10^{-5}$, the recompression phenomenon always occurs when the incident shock Mach number is larger than 1.4. However, it is also sensitive to droplet volume fraction. For an incident shock Mach of 1.6, recompression only can occur with initial droplet volume fractions below $3.28 \times 10^{-5}$.

**Acknowledgement**

The computational work for this article was fully performed on resources of the National Supercomputing Center, Singapore (https://www.nscc.sg/). This work is supported by Singapore

Ministry of Education Tier 1 Grant (R-265-000-688-114).

**Data availability**

The data that support the findings of this study are available from the corresponding author upon reasonable request.

(2019).